\documentclass[a4paper,10pt,twocolumn,twoside]{article}

\usepackage{amssymb,amsfonts,amsmath,mathtext,float,geometry}
\usepackage{subfigure}

\usepackage{graphicx}% Include figure files

\usepackage{enumitem} %% to adjust the enumerate environment.
\usepackage[usenames,dvipsnames,svgnames,table]{xcolor}

\newcommand{\myToC}[3]{\item[\ref{#3}.]  #1  \small  \textit{#2} \normalsize   \dotfill  \pageref{#3}} %%% REMOVE ME
\newcommand{\mysubToC}[2]{\item[--] \textit{#1}  \dotfill  \pageref{#2}} %%% REMOVE ME

\usepackage{fancybox,fancyhdr,lastpage,}%this packages provides fancy up and bottom of page
\fancypagestyle{plain}{%
\fancyhead[R]{}
\fancyhead[L]{}
\fancyhead[C]{}

\fancyfoot[R]{\small \textit{M. Konnik and J.Welsh} \normalsize}
\fancyfoot[L]{\small \textit{High-level noise simulations in CCD/CMOS} \normalsize}
\fancyfoot[C]{ \textit{page \thepage \,of \pageref{LastPage}.} }
}

\title{\textbf{High-level numerical simulations of noise in CCD and CMOS photosensors:  review and tutorial}}
\author{Mikhail Konnik\footnote{Corresponding author. Emails: \textit{mikhail.konnik@uon.edu.au} or \textit{mail@mvkonnik.info} } \,\, and James Welsh}
\date{Faculty of Engineering and Built Environment, University of Newcastle, Australia}
\begin{document}
\maketitle

\begin{abstract}
In many applications, such as development and testing of image processing algorithms, it is often necessary to simulate images containing realistic noise from solid-state photosensors. A high-level model of CCD and CMOS photosensors based on a literature review is formulated in this paper. The model includes photo-response non-uniformity, photon shot noise, dark current Fixed Pattern Noise, dark current shot noise, offset Fixed Pattern Noise, source follower noise, sense node reset noise, and quantisation noise. The model also includes voltage-to-voltage, voltage-to-electrons, and analogue-to-digital converter non-linearities. The formulated model can be used to create synthetic images for testing and validation of image processing algorithms in the presence of realistic images noise. An example of the simulated CMOS photosensor and a comparison with a custom-made CMOS hardware sensor is presented. Procedures for characterisation from both light and dark noises are described. Experimental results that confirm the validity of the numerical model are provided. The paper addresses the issue of the lack of comprehensive high-level photosensor models that enable engineers to simulate realistic effects of noise on the images obtained from solid-state photosensors.
\end{abstract}

\pagestyle{plain}

\section{Introduction}\label{sec:introduction}
Many modern measuring devices use CMOS or CCD solid-state photosensors to convert light into a digital signal. Due to imperfections of photosensors such a conversion is not ideal and leads to noise in the measured signal. Therefore, one can either estimate and reduce the impact of noise from the sensor, or simulate and predict the impact of noise on the images quality. The problem of noise estimation has been addressed by the EMVA1288 standard~\cite{EMVA1288}. The high-level simulation of noise in photosensors, however, is still an area of active research. The main problem is that photosensors are affected by many different sources of noise, some of which cannot be modelled adequately using only Gaussian noise.

\label{review:2R2}Different numerical models of CMOS~\cite{kolehmainen2004simulation,wach2004noise} and CCD~\cite{boiecoxccdnoisemodel,costantini2004virtual,jiang1997detecting} sensors are reported in the literature. The photosensor models differ by the generality and the scope of the noise sources being included. High-level  simulators~\cite{costantini2004virtual} are used for the evaluation of how different noise sources influence image quality~\cite{farrell2004simulation}, while low-level simulator models specific aspects of the solid-sate physics of a sensor ~\cite{shcherback2003photoresponse}. The thoroughness of noise modelling differs as well; for example, in some models of the photosensor, the dark current shot noise is described by its mean value~\cite{donaghue1993model,mullikin1994methods}, while other models use more sophisticated statistical description of noise~\cite{baer2006model}. \label{review2JEI:R1Q1-2} Sophisticated  software integrated suits such as the Image Systems Evaluation Toolkit~\cite{farrell2008sensor,farrell2004simulation,vora2001image} (ISET) were developed recently.

\paragraph*{Models of CMOS photosensors:}
\label{review:1R2} Both high-level~\cite{farrell2004simulation,kolehmainen2004simulation,wach2004noise,costantini2004virtual,1fnoiseincmoselgamal,holstccdarrayscamerasdisplays} and low-level~\cite{apscrosstalks,qemodellinghornsey} numerical models of CMOS photosensors are reported in the literature. Low-level simulators like Medici~\cite{medicicmossimulator}, which are able to model the 2-D distributions of potential and carrier concentrations, are used for simulation of solid-state physics in CMOS photosensors~\cite{hornseyCMOSanalysis2}. Some of the papers deal with the numerical modelling of colour images that allow end users to see the effects of the noise on the images~\cite{farrell2008sensor}. Other papers focus on the specific types of noise; for example, the analysis of temporal noise in CMOS active pixels sensors can be found in~\cite{tian2002analysis}, an autoregressive model of fixed-pattern noise (FPN) was presented in~\cite{elgamalFPNmodeling} along with other FPN models~\cite{kelly2008fixed,fowler1998method}. High-level models (e.g.,~\cite{comprehcmosmodel,irie2008model}) often assume that the ADC, sense node and source follower are completely linear, which, as we will show in this paper, is not always the case.

\paragraph*{Models of CCD photosensors:}
The models of CCD photosensors~\cite{irie2008model,irieccdnoiseevaluation} and their noise parameters estimation~\cite{irieccdnoiseevaluation} are generally simpler than models of CMOS sensors. The model by Flory~\cite{floryimageacquisition} describes the noise of the sensor electronics as a combination of shot noise and amplifiers noise. The CCD camera noise model by Cox~\cite{boiecoxccdnoisemodel} includes photon and electronic shot noise, dark-current noise and readout noise; however the model suggests that noise is stationary. Healey and Kondepudy~\cite{Healeyradiometricccd} model camera noise with offset fixed-pattern noise (FPN), photon and dark-current shot noise, read noise and photo response non-uniformity (PRNU). \label{review2JEI:R2_Q2}An adequate model was proposed by Costantini and S\"{u}sstrunk~\cite{costantini2004virtual}, where noises was categorised in four main types: photon shot noise, Photo Response Non Uniformity, dark current shot noise and read noise.

\paragraph*{The purpose of the article:}
\label{review:18R2} This article provides a literature review of the noise models that are used in the simulations of CCD and CMOS photosensors and give a working numerical model of a photosensor. The aim of this article is to summarise the advances in numerical simulations of photosensors and provide a MATLAB implementation of a CCD/CMOS sensor model, which has been demonstrated to adequately describe noise properties of a hardware CMOS photosensor.

The experimental results for a hardware 5T CMOS photosensor are provided in the article as validation for the developed numerical simulator. The article addresses the issue of the lack of high-level models of photosensors. It is shown that the dark noise may have more complicated structure than is typically assumed and therefore requires a more sophisticated statistical description. The formulated high-level model of a photosensor allows the simulation of realistic noise effects on the images and can be useful for testing of image processing algorithms.

\paragraph*{Software implementation:}
The MATLAB-based software implementation of the high-level CCD/CMOS photosensor model described in this article is freely available on the website \verb|https://code.google.com/p/highlevelsensorsim/| and also on email request. The software simulator written as a series of functions that call the corresponding sub-functions sequentially to add the noise with specified parameters. Two example test files are supplied that run: (1) a \textbf{simple} sensor model, which is completely linear and where all noise are Gaussian, and (2) an \textbf{advanced} model, which has V/V and V/e non-linearities, LogNormal noise models and other specific noises. Both test files use the same code:  adjusting the parameters in the test file, a user can turn the noise sources on or off and trigger non-linearities.  The MATLAB code has a thorough documentation, both as comments inside the m-files and rendered for web view~\cite{konnikCodemanagement2012} in a browser (see \textit{/help} directory and \textit{index.html} file inside).

\subsection*{Contents of the article}\label{sec:contents}

\begin{enumerate}[topsep=0ex,itemsep=0ex,partopsep=1ex,parsep=1ex,leftmargin=2ex]

\myToC{Introduction}{}{sec:introduction}

\myToC{Noise sources in solid-state photosensors:}{}{sec:noisecontributions} 

	\begin{enumerate}[topsep=0ex,itemsep=0ex,partopsep=0ex,parsep=1ex]
	 \myToC{From Photon to Charge:}{}{sec:photon2electron}

		\begin{enumerate}[topsep=-2ex,itemsep=0ex,partopsep=0ex,parsep=0ex,leftmargin=0ex]
		\mysubToC{photon shot noise and PRNU}{sec:prnumodel}
		\mysubToC{dark signal, dark current FPN and dark current shot noise}{subsec:darkcurrentnoise}
		\mysubToC{Source follower noise (Johnson noise, Flicker (1/f) noise, and  Random Telegraph Noise)}{sec:SFnoise}
		\end{enumerate}

	 \myToC{From Charge to Voltage:}{}{sec:electron2voltage}

		\begin{enumerate}[topsep=-2ex,itemsep=0ex,partopsep=0ex,parsep=0ex,leftmargin=0ex]
		\mysubToC{Sense node reset noise (kTC noise)}{sec:subsection:snresetnoise}

		\mysubToC{Offset FPN}{subsec:offsetFPNdescription}

		\mysubToC{Gain non-linearities (V/$e^-$ gain and  V/V gain non-linearity) }{sec:vvnonlin}

		\mysubToC{Correlated double sampling}{subsec:cdsdescription}
		\end{enumerate}

	 \myToC{From Voltage to Digital Numbers}{(including Differential and Integral Linearity Errors)}{sec:adc}

	\end{enumerate}

\myToC{Simulation Methodology}{}{sec:simulationmethodology}

	\begin{enumerate}[topsep=0ex,itemsep=0ex,partopsep=0ex,parsep=1ex]

	 \myToC{Model: From Photon to Charge:}{}{seq:photon2charge}

		\begin{enumerate}[topsep=-2ex,itemsep=0ex,partopsep=0ex,parsep=0ex,leftmargin=0ex]
		\mysubToC{simulating photon shot noise and PRNU}{subseq:photonshotsimulation}
		\mysubToC{converting photons to electrons}{subsec:convertingPhToE}
		\mysubToC{simulating dark signal, dark current FPN and dark current shot noise}{subsec:darkcurrentshotnoisesimulation}
		\mysubToC{simulating Source follower noise}{sec:sf:subsec:sfnoisesimulate}
		\end{enumerate}

	 \myToC{Model: From Charge to Voltage:}{}{sec:chargevoltagesimulation}

		\begin{enumerate}[topsep=-2ex,itemsep=0ex,partopsep=0ex,parsep=0ex,leftmargin=0ex]
		\mysubToC{simulating sense node reset noise (kTC noise) and offset FPN}{sec:subses:elgamalfpn}
		\mysubToC{converting electrons to voltage}{sec:simulationSourceFollower}
		\mysubToC{simulating  V/$e^-$  and  V/V gain non-linearity}{subses:vvnonlinsimulation}
		\end{enumerate}

	 \myToC{Model: From Voltage to Digital Numbers}{}{sec:fromvoltagetodn}

		\begin{enumerate}[topsep=-2ex,itemsep=0ex,partopsep=0ex,parsep=0ex,leftmargin=0ex]
		\mysubToC{simulating ADC non-linearity}{subsec:adcnonlinandnoise}
		\mysubToC{simulating quantisation noise}{subsec:adcnoise}
		\end{enumerate}

	\end{enumerate}

\myToC{Experimental validation of the photosensor model}{}{sec:experimentalvalidation}

	\begin{enumerate}[topsep=0ex,itemsep=0ex,partopsep=0ex,parsep=1ex]

	\myToC{Radiometric function}{}{subsec:radiometricfunc}
	\myToC{Photon Transfer Curve}{}{subsec:ptcsim}
	\myToC{Signal-To-Noise Ratio}{}{subsec:snrmeasurements}
	\myToC{Photo Response Non-Uniformity}{}{subsec:PRNU}
	\myToC{Noise Spectrogram}{}{subsec:spectrogram}
	\myToC{Dark signal performance for different integration times:}{}{subsec:darkfpnperf}

		\begin{enumerate}[topsep=-2ex,itemsep=0ex,partopsep=0ex,parsep=0ex,leftmargin=0ex]
			\mysubToC{Short integration time}{sec:simulations:subsubsec:shorttime}
			\mysubToC{Long integration time}{subsec:verylongfpn}
		\end{enumerate}

	\end{enumerate}

\myToC{Conclusion}{}{sec:conclusion}

\end{enumerate}

\section{Noise sources in solid-state photosensors}\label{sec:noisecontributions}
Many noise sources contribute to the resulting noise image that is produced by photosensors. Noise sources can be broadly classified as either \textbf{fixed-pattern} (time-invariant) or \textbf{temporal} (time-variant) noise. Fixed-pattern noise refers to any spatial pattern that does not change significantly from frame to frame. Temporal noise, on the other hand, changes from one frame to the next. These noise sources will be described in detail in the following subsections.  \label{review2JEI:R2_Q3}

\subsection{From Photon to Charge}\label{sec:photon2electron}
The ability of a semiconductor to produce electrons from incident photons is referred to as \textit{quantum efficiency} (QE). \label{review:5R2}\label{review:29R1} Quantum efficiency can be supplied by the sensor manufacturer in the form of spectral responsivity as a function of wavelength, or measured empirically. \label{review2JEI:R2Q9-1}For the sake of simplicty, we assume that the light is monochromatic, and therefore we need the spectral response of the sensor at one wavelength.

The charge collected in each pixel of a sensor array is converted to voltage by a \textit{sense capacitor} and a \textit{source-follower amplifier}. The sense node is shared for all pixels in the case of a CCD. In CMOS sensors, the sense node is located inside each pixel.

\subsubsection{Photon shot noise}\label{sec:photonshot}
The process of photon capturing has an uncertainty that arises from random fluctuations when photons are collected by the photodiode. Such uncertainty leads to \textit{photon shot noise} and is described by the Poisson process. In the case of a high level  of light  (more than 1000 arrival events~\cite{mendenhallstatistics} of photons), the Poisson distribution may be approximated by a Gaussian distribution. However, this is not justified for low-light conditions. The method of the approximation of the Poisson distribution by Gaussian is discussed in detail in~\cite{poisson2gaussapprox}.

\subsubsection{Photo response non-uniformity}\label{sec:prnumodel}
\label{review:12R1}The Photo Response Non-Uniformity (PRNU) is the spatial variation in pixel output under uniform illumination mainly due to variations in the surface area of the photodiodes. This occurs due to variations in substrate material during the fabrication of the sensor~\cite{hornseynoisebook}. 

The PRNU is signal-dependent (proportional to the input signal) and is fixed-pattern (time-invariant). \label{review:30R1} The PRNU factor is typically $0.01\dots 0.02$ for a given sensor~\cite{janesickscintificCCD}, but varies from one sensor to another. It is noteworthy that the PRNU pattern can be used as a unique identification fingerprint for digital cameras~\cite{prnucameraID}.

\subsubsection{Read noise}\label{sec:readnoise}
\label{review:13R1}
Read noise is defined as any noise that is not a function of the signal~\cite{photontransferbook}; it is a combination of the remaining noise generated between the photodiode and the output of the ADC circuitry. The read noise consists of: \textit{dark current shot noise} (Sec.~\ref{subsec:dshot}), \textit{dark current Fixed Pattern Noise} (Sec.~\ref{subsec:DFPN}),  \textit{sense node reset noise} (Sec.~\ref{sec:subsection:snresetnoise}), \textit{source follower noise} (Sec.~\ref{sec:SFnoise}), \textit{ADC quantisation noise} (Sec.~\ref{subsec:adcnoise}).

\subsubsection{Dark current}\label{subsec:darkcurrentnoise}
Dark current is the thermally generated electrons~\cite{prnuqualification} discharging the pixel just as if a photon had hit the pixel. The dark current generally arises from surface defects in the $SiO_2/Si$ interface~\cite{theuwissen1995solid}, surface generation~\cite{leakagemodellingCMOS}, thermal generation~\cite{shockley1952statistics}, and imperfections of the semiconductor manufacturing process~\cite{takayanagi2003dark}. 

The average dark current $D_R$ [e$^-$/sec/pixel] can be characterised by~\cite{janesickscintificCCD}:\label{review:16R1}
\begin{equation}\label{eq:darkcurrentrate}
 D_R = P_A D_{FM} T^{3/2} \exp\left(-\frac{E_{gap}}{2\cdot k T}\right),
\end{equation}
where $P_A$ is the pixel area [cm$^2$], $D_{FM}$ [nA/cm$^2$] is the dark current \label{review:7R2}figure-of-merit\footnote{The dark current figure-of-merit $D_{FM}$ can be estimated from the measurements to ensure that the number that manufacturers publish is consistent with the sensor performance. Using Photon Transfer methodology~\cite{photontransferbook}, one can estimate the dark current FPN factor $D_N$ and the dark current figure-of-merit $D_{FM}$. This can be done by plotting dark current shot noise and dark current FPN versus a dark   signal. This is called ``Dark Transfer Curve'', out of which those parameters can be deduced. The procedure is described on page 170-171 in~\cite{photontransferbook}.} (varies with the sensor and usually reported by the manufacturer) at 300K, $E_{gap}$ [eV] is the band gap energy of the semiconductor which also varies with temperature, $T$ is the temperature in K, and $k$ is Boltzman's constant.

\label{review:37R1}It was reported~\cite{widenhorn2002temperature} that for a CCD photosensor operated in multipinned phase (MPP) mode, the  average dark current $D_R $ can be described as:
\begin{equation}\label{eq:darkcurrentrateother}
 D_R \sim T^3 \exp\left(-\frac{E_{gap}}{kT}\right) + T^{3/2} \exp\left(-\frac{E_{gap}}{2kT}\right),
\end{equation}
where the first term is more important as the temperature increases, whereas the second term dominates at lower temperatures. Temperature dependence of the dark current is explained in details in~\cite{szesolidstatephysics,widenhorn2001meyer,widenhorn2002temperature}.

The relationship between band gap energy and temperature can be described by Varshni's empirical expression~\cite{Pinault20011562}:
\begin{equation}\label{eq:darkcurrentegap}
     E_{gap}(T)=E_{gap}(0)-\frac{\alpha T^2}{T+\beta},
\end{equation}
where $E_{gap}(0)$, $\alpha$ and $\beta$ are material constants. For Silicon $E_{gap}(0) = 1.1557 [eV]$,  $\alpha = 7.021*10^{-4}$ [eV/K], and $\beta = 1108$ [K].

\subsubsection{Dark signal}
\label{review:15R1}The dark signal varies from pixel to pixel, it is linear with respect to integration time, and doubles with every $6-8^{\circ}C$ increase of temperature~\cite{hornseynoisebook}. When the exposure begins, a dark current is generated even if there is no light. The longer the exposure time $t_I$, the more dark   signal $S_{dark.e^-}$ (number of electrons per pixel) will be generated:
\begin{equation}\label{eq:darkcurrent}
S_{dark.e^-} = t_I\cdot D_R,
\end{equation}
\label{review:13-2R1}where $D_R$ is the average dark current from Eq.~\ref{eq:darkcurrentrate}. In simulation we use Eq.~\ref{eq:darkcurrent} to calculate the dark   signal per pixel. But the dark   signal $S_{dark.e^-}$  is a subject of the \textit{dark current shot noise} due to random generation of the electrons.

\subsubsection{Dark current shot noise}\label{subsec:dshot}
\label{review:37-1R1} An additional noise results from the dark   signal due to the electrons   being generated randomly by the photosensor. Such noise is called dark current shot noise~\cite{janesickscintificCCD} and is described by the Poisson distribution as the random arrival of electrons~\cite{darknoisereductionPLAD} as:
\begin{equation}\label{eq:darkshotnoise}
 \sigma_{d.shot} = \sqrt{t_I\cdot D_R} = \sqrt{S_{dark.e^-}},
\end{equation}
where $D_R$ is the  average dark current [$e^-$/sec/pixel] as described by Eq.~\ref{eq:darkcurrentrate}. In a real sensor, the pixels differ slightly from one to another \label{review2JEI:R2Q5} resulting in another source of noise called \textit{dark current Fixed Pattern Noise}.

\subsubsection{Dark current fixed pattern noise}\label{subsec:DFPN}
\label{review:33R1}Pixels in a hardware photosensor cannot be manufactured exactly the same from perfectly pure materials. There will always be variations in the photo detector area that are spatially uncorrelated~\cite{ccduncorrelated}, surface defects at the $SiO_2/Si$ interface~\cite{sakaguchidarkcurrentreduction}, and discrete randomly-distributed charge generation centres~\cite{baer2006model}. These defects provide a mechanism for thermally-excited carriers to move between the valence and   conduction bands~\cite{mcgrath2005counting,hawkinsgenerationcurrents}. Consequently, the average dark   signal is not uniform but has a spatially-random and fixed-pattern noise (FPN) structure.  The dark current FPN can be expressed ~\cite{photontransferbook} as follows:\label{review:32R1}
\begin{equation}\label{eq:darkcurrentFPN}
 \sigma_{d.FPN} = t_I D_R \cdot D_N,
\end{equation}
where $t_I$ is the integration time, $D_R$ is the  average dark current described in Eq.\ref{eq:darkcurrentrate}, and $D_N$ is the dark current FPN factor that is typically $0.1\dots 0.4$ for CCD and CMOS sensors. 

There are also so called ``outliers'' or ``dark spikes''~\cite{janesick2010fundamental}; that is, some pixels generate a dark signal values much higher than the mean value of the dark signal. The mechanism of such ``dark spikes'' or ``outliers'' can be described by the Poole-Frenkel effect~\cite{Harrell1999195,polefrenkeleffect} (increase in emission rate from a defect in the presence of an electric field).

\subsubsection{Source follower noise}\label{sec:SFnoise}
\label{review:4R1}In high-end CCD and CMOS sensors the source follower noise has been decreased to a value of one electron rms~\cite{photontransferbook}. However, source follower noise for industry-grade sensors can be significant and therefore should be included in a photosensor model. The source follower noise consists of  white noise,  flicker noise (1/f noise), and  random telegraph noise (RTN).

\paragraph{Johnson noise (white noise)}
Similar to the sense node, the source follower amplifier has a resistance that generates thermal noise. Such noise is governed by the Johnson white noise equation~\cite{hamamatsuflicker}. The noise is commonly referred to as Johnson noise, Johnson-Nyquist or simply as a white noise~\cite{hornseynoisebook}.

\paragraph{Flicker (1/f) noise}
The flicker noise, also referred to as 1/f noise, is generally due to imperfect contacts between two materials at the junction~\cite{ott1988noise,nakamura2006image}, including metal-to-metal, metal-to-semiconductor and semiconductor-to-semiconductor.  Since MOSFETs are used inside each pixel, CMOS image sensors exhibit higher 1/f noise than CCD sensors. More details and discussions about 1/f noise can be found in~\cite{1fnoiseincmoselgamal,tian1fanalysis}.

\paragraph{Random Telegraph Noise}
As the CCD and CMOS pixels are shrinking~\cite{rhodes2005cmos,schoberl2012photometric} in dimension, the low-frequency noise is subsequently increasing~\cite{kolhatkar2004separation}. The origin of RTN is attributed to the random trapping and emission of mobile charge carriers resulting in discrete modulation of the channel current, which can be modelled~\cite{1fnoiseincmoselgamal} as a random telegraph signal. However, the modelling and explanation of both flicker noise and RTN are still subjects of active research~\cite{antal20011,bogaerts2002rts,hopkins1993}. It was shown~\cite{antal20011} that voltage fluctuations, which exhibit a $1/f$ power spectrum, can be described by the Fisher-Tippet-Gumbel distribution. Further details about RTN noise in photosensors can be found in~\cite{bogaerts2002rts,hopkins1993}.

\subsection{From Charge to Voltage}\label{sec:electron2voltage}
The conversion from charge to voltage is not perfect in the real photosensor: sense node reset noise, source follower noise, and offset Fixed Pattern Noise add an uncertainty to the measured signal~\cite{tian2002analysis}. These noise sources are described in this subsection.

\subsubsection{Sense node reset noise (kTC noise)}\label{sec:subsection:snresetnoise}
Prior to the measurement of each pixel's charge packet, the sense node capacitor is reset to a reference voltage level~\cite{hamamatsuflicker}. Noise is generated at the sense node by an uncertainty in the reference voltage level due to thermal variations in the channel resistance of the MOSFET reset transistor. The reference level of the sense capacitor is therefore different from pixel to pixel~\cite{hamamatsuflicker}. It was reported~\cite{fowler2000low} that the reset noise can be a significant contributor to dark noise. The reset noise voltage is given by:
\begin{equation}\label{eq:ktcnoise}
\sigma_{RESET}  = \sqrt{\frac{k_B T}{C_{SN}}},
\end{equation}
where $k_B$ is Boltzmann's constant, $C_{SN}$ is the sense node capacitance [F], and $T$ is the  temperature [K].

Because reset noise can be significant~\cite{hamamatsuflicker} (about 50 rms electrons), most high-performance photosensors incorporate a noise-reduction mechanism such as correlated double sampling (CDS; see subsection \ref{subsec:cdsdescription} and subsection \ref{subsec:CDSsimulation}).

\paragraph*{Sense node reset noise for CCD and CMOS sensors:}
For CCD sensors, the sense node reset noise is removed by CDS~\cite{photontransferbook}. In CMOS photosensors, it is difficult to remove the reset noise given the specific CMOS pixel architectures, even after application of CDS. The CDS suppresses the low frequency noise components, although it increases the thermal noise contributions~\cite{chargetracker}.

\subsubsection{Offset fixed pattern noise}\label{subsec:offsetFPNdescription}
Pixels in the same column of the photosensor share a column amplifier. \label{review:36-1R1}Differences in the gain and offset of these column amplifiers contribute to a column-wise offset fixed pattern noise. The offset fixed pattern noise (offset FPN) is caused by an offset in the integrating amplifier, size variations in the integrating capacitor, and as variation of bias/offset voltages~\cite{elgamalFPNmodeling}. 

\label{review:6R2} Specifically, CMOS offset FPN is due to threshold voltages differences in the pixel source follower amplifier~\cite{janesickpersonal}. Offset FPN arises from the fact that pixels in the same column of a CMOS sensor share the same column amplifier. The difference in gain and the offsets of such an amplifier is a source of column-to-column offset FPN ~\cite{elgamalFPNmodeling,kelly2008fixed}. This type of noise appears as ``stripes'' in the image and can result in significant image quality degradation. Modelling of the offset FPN is discussed in Subsection~\ref{sec:subses:elgamalfpn}.

\subsubsection{Gain non-linearity in photosensors}\label{sec:vvnonlin}
Both CCD and CMOS sensors may exhibit V/V (voltage-voltage conversion) \label{review:8R2} and V/$e^-$ (voltage-electrons conversion) non-linearity. The V/V non-linearities is mainly~\cite{janesickscintificCCD} due to the source follower amplifier. The V/$e^-$ non-linearity is due to the sense node capacitance $C_{SN}$. The V/$e^-$ non-linearity is a problem mostly for CMOS sensors, whereas for   CCD sensors the V/$e^-$ non-linearity is usually negligible.

\subsubsection{V/$e^-$ gain non-linearity}
The V/$e^-$ non-linearity affects both FPN and shot noise. It can also be thought of as a sense node capacitor non-linearity. When a small signal is measured, the sense node capacitance $C_{SN}$ may change negligibly. For a large signal, a change of $C_{SN}$ affects the signal being measured. The V/$e^-$ non-linearity can be expressed as~\cite{photontransferbook}:\label{review:17R1}
\begin{equation}\label{eq:venonlinsignal}
 S_{e^-} = \frac{k_1}{q} \ln \left[ \frac{V_{REF}}{ V_{SN} } \right],
\end{equation}
where $S_{e^-}$ is the total signal in electrons, the constant $k_1$ is $k_1=10.909*10^{-15}$, $V_{REF}$ is the reference voltage to which the sense node is reset.

\subsubsection{V/V gain non-linearity}\label{sec:nonlin:subsec:vvnonlin}
The V/V non-linearity is caused by the non-linear response of source follower amplifier~\cite{photontransferbook}. In our simulations, we implemented the linear change of source follower gain $A_{SF}$ similar to the model in~\cite{photontransferbook}. The V/V gain non-linearity is described as:
\begin{equation}\label{eq:sfnewgain}
A_{SF_{new}} = (\gamma_{nlr} -1) \cdot \frac{V_{SN}}{V_{FW} } + A_{SF},
\end{equation}
\label{review:17-1R1}where $V_{SN}$ is the voltage that corresponds to the signal collected by the sense node, $V_{FW}$ is the voltage that corresponds to the full-well signal, and $\gamma_{nlr}$ is used to control the amount of non-linearity in the source follower gain $A_{SF}$. The parameter of non-linearity $\gamma_{nlr}$ can be taken from the specifications of a photosensor. The new source follower gain $A_{SF_{new}}$  from Eq.~\ref{eq:sfnewgain} is then used for the conversion in Eq.~\ref{eq:conversiontosourcefollower} in Subsection~\ref{sec:simulationSourceFollower} to simulate the V/V non-linearity.

\subsubsection{Correlated double sampling}\label{subsec:cdsdescription}
\label{review:3R1}Fixed pattern noise performance of CMOS sensors is usually lower than for CCD sensors~\cite{fossumcameraonachip}. For this reason CMOS sensors use noise reduction circuits such as correlated double sampling (CDS). The  CDS circuits are located per column and usually consist of two sample-and-hold (S\&H) circuits~\cite{mendis1997cmos}. During the pixel read-out cycle, two samples are taken: the first when the pixel is in the reset state and the second   when the charge has been transferred to the read-out node.

During the reset stage, the photodiode capacitance is charged to a reset voltage. The reset voltage is read by the first sample-and-hold (S\&H) in a correlated double sampling (CDS) circuit~\cite{correlateddoublesampling}.  Then the exposure begins: the photodiode capacitor is discharged during an exposure (integration) time at a rate proportional to the incident illumination. This voltage is then read by the second sample-and-hold of the CDS. The CDS circuit subtracts the signal pixel value from the reset value. 

The main purpose of CDS is to eliminate fixed pattern noise caused by random variations in the threshold voltage of the reset and pixel amplifier transistors. Many variants of CDS were proposed and implemented~\cite{white1974characterization,xu2002new,koklu2011switched}. Many sensors use correlated double sampling to eliminate column patters and pixels patterns. Pixel noise reduced by the CDS is known as ``crowbar'' that significantly reduces the effect of the reset column noise~\cite{mendis1997cmos,blanksby2002performance}.

\subsection{From Voltage to Digital Numbers}\label{sec:adc}
\label{review:34R1}An analogue-to-digital converter (ADC) transforms a voltage signal into discrete codes. 
An $N$-bit ADC has $2^N$ possible output codes with the difference between code being $V_{ADC.REF}/2^N$. The resolution of the ADC indicates the number of discrete values that can be produced over the range of analogue values and can be expressed as:
\begin{equation}\label{eq:kadc}
 K_{ADC} = \frac{V_{ADC.REF} - V_\mathrm {min}}{N_{max}}
\end{equation}
where $V_\mathrm{ADC.REF}$ is the maximum voltage that can be quantified, $V_\mathrm {min}$ is minimum quantifiable voltage, and $N_{max} = 2^N$ is the number of voltage intervals. Therefore, the output of an ADC can be represented as:
\begin{equation}\label{eq:adcconverter}
ADC_ \mathrm {Code} = \textrm{round}\left( \frac{V_\mathrm {input}-V_ \mathrm {min}}{K_{ADC}} \right)
\end{equation}
The lower the reference voltage $V_{ADC.REF}$, the smaller the range of the voltages one can measure.

\subsubsection{Noise and non-linearity induced by an ADC}\label{sec:adcnonlin}
In terms of the ADC, the following non-linearity and noise should be considered for the simulations of the photosensors: Integral Linearity Error, Differential Linearity Error, quantisation error, and ADC offset.

\paragraph*{Differential Linearity Error } (DLE) indicates the deviation from the ideal 1 LSB (Least Significant Bit) step size of the analogue input signal corresponding to a code-to-code increment~\cite{razavi1995principles}. Assume that the voltage that corresponds to a step of 1 LSB is $V_{LSB}$. In the ideal case, a change in the input voltage of $V_{LSB}$ causes a change in the digital code of 1 LSB. If an input voltage that is more than $V_{LSB}$ is required to change a digital code by 1 LSB, then the ADC has \textit{DLE error}\label{review:35R1}. In this case, the digital output remains constant when the input voltage changes from, for example, $2 V_{LSB}$  to  $4 V_{LSB}$, therefore corresponding the digital code can never appear at the output. That is, that code \textit{is missing}.

\paragraph*{Integral Linearity Error} (ILE) is the maximum deviation of the input/output characteristic from a straight line passed through its end points~\cite{razavi1995principles}. For each voltage in the ADC input, there is a corresponding code at the ADC output. If an ADC transfer function is ideal, the steps are perfectly superimposed on a line. However, most real ADC's exhibit deviation from the straight line, which can be expressed in percentage of the reference voltage or in LSBs. Therefore, ILE is a measure of the straightness of the transfer function and can be greater than the differential non-linearity. Taking the ILE into account is important because it cannot be calibrated out. 

\paragraph*{Quantisation errors} are caused by the rounding, since an ADC has a finite precision. The probability distribution of quantisation noise is generally assumed to be uniform. Hence we use the uniform distribution to model the rounding errors (see Subsection~\ref{subsec:adcnoise}).

\paragraph*{ADC offset error} may occur due to the DC offset associated with the analogue inputs to the ADC. The magnitude of the ADC offset depends on the gain and input range selection.

\section{Simulation Methodology}\label{sec:simulationmethodology}
\begin{figure}[ht!]
\centering\includegraphics[width=6.45cm]{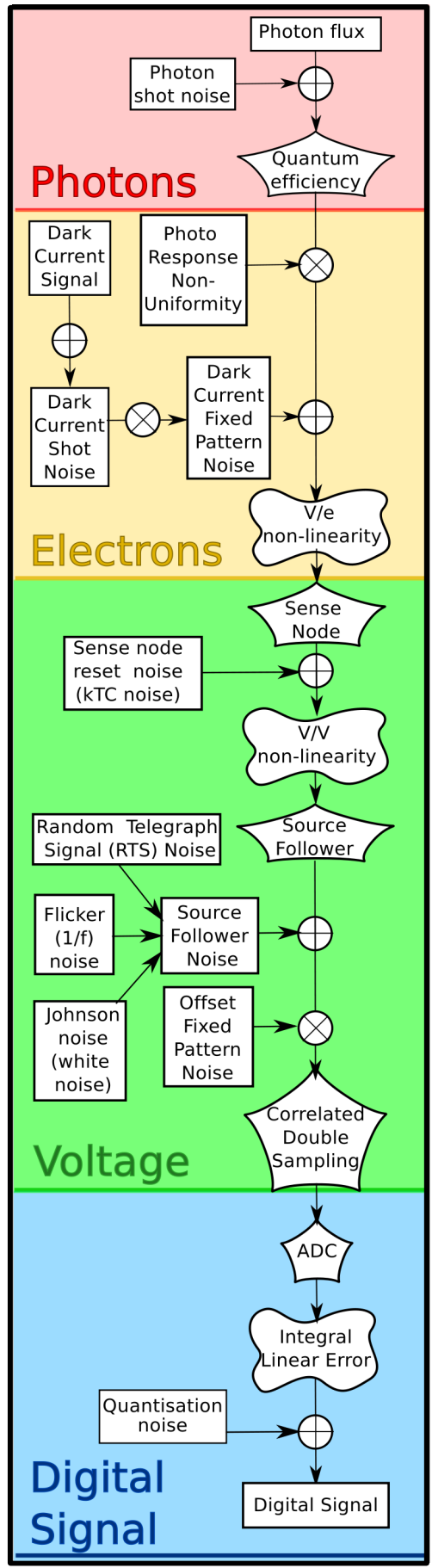} %%% this is for FINAL VERSION
\caption{The diagram of the photosensor model.}
\label{fig:camerascheme}
\end{figure}

The photosensor model was implemented in MATLAB as a set of functions that generate noise according to the models discussed above. The functions sequentially transform the input to number of photons, to number of electrons, to voltages, and finally to a digital signal, as shown in Fig.~\ref{fig:camerascheme}. One can simulate either CCD or CMOS photosensors by turning the models of noise on and off. The simulation software contains routines for the simulation of the sensor and also scripts for the estimation of the noise characteristics. 

~\label{review:11-diagramR1}

\label{review:21-8R1} The numerical model of the photosensor simulates the temporal and fixed-pattern characteristics of noise sources accordingly. The matrices for the fixed-pattern (time-invariant) noise are calculated once and then saved to a file. Those matrices will be loaded again to calculate the fixed pattern noise; therefore, the noise will always remain the same reflecting the fixed pattern nature of the noise. On the other hand, the matrices for the temporal (time-variant) noise are recalculated each time the simulation begins.

\subsection{Model: From Photon to Charge}\label{seq:photon2charge}
\label{review:5R1} The input to the model of the photosensor is assumed to be a matrix $U_{in} \in \mathbb{C}^{N\times M} $ that corresponds to the opical field. For the estimation purpose, the input  can be a uniform light field (i.e., uniform image) or \label{review:5-1R1} gradient image (i.e., linear transition from no light to saturation level). 
\label{review2JEI:R1_Q3} Then the sensor's irradiance $I_{irrad} = |U_{in}|^2$, which is in units of  [W/m$^2$], converted to the average number of photons $I_{ph}$ collected by each pixel during the integration (exposure) time:
\begin{equation}\label{eq:photonsignal}
 I_{ph}  =  round \left( \frac{ I_{irrad} \cdot P_A  \cdot t_I }{ Q_p} \right),
\end{equation}
where $P_A$ is the area of a pixel [m$^2$],  $t_{I}$ is integration (exposure) time, $Q_p = \frac{h\cdot c}{\lambda}$ is the energy of a single photon at wavelength $\lambda$, $h$ is Planck's constant and $c$ is the speed of light. 

Therefore, each $(i,j)$-th element of the \label{review:21-1R1} matrix $I_{ph}$ contains the number of photons that have been collected by the $(i,j)$-th pixel of the ``software'' photosensor during the integration time $t_{I}$.

\subsubsection{Simulating photon shot noise}\label{subseq:photonshotsimulation}
The photon shot noise is due to the random arrival of photons and can be described by a Poisson process as discussed in Subsection~\ref{sec:photonshot}. \label{review:20R1}Therefore, for each $(i,j)$-th element of the matrix $I_{ph}$ that contains the number of collected photons, a photon shot noise  is simulated as a Poisson process $\mathcal{P}$   with mean $\Lambda$:
\begin{equation}\label{eq:photonshotnoisesimulations}
 I_{ph.shot}  = \mathcal{P}(\Lambda), \,\,\,\,\mbox{ where   } \Lambda = I_{ph} .
\end{equation}
In MATLAB, we   use the \texttt{poissrnd} function that generates Poisson random numbers with mean $\Lambda$.  \label{review:23R1} That is, the number of collected photons in \label{review:21R1} $(i,j)$-th pixel of the simulated photosensor in the matrix $I_{ph}$ is used as the mean $\Lambda$ for the generation of Poisson random numbers to simulate the photon shot noise. The input of the \texttt{poissrnd} function will be the matrix $I_{ph}$ that contains the number of collected photons. The output will be the matrix $I_{ph.shot} \rightarrow I_{ph}$, i.e., the signal  with added photon shot noise.  \label{review:21-7R1} The matrix $I_{ph.shot}$ is recalculated each time the simulations are started, which corresponds to the temporal nature of the photon shot noise.

\subsubsection{Conversion to Electrons}\label{subsec:convertingPhToE}
\label{review:9R2}
The matrix that contains the number of collected photons during the exposure time, including the  photon shot noise $I_{ph}$, is converted to the matrix $I_{e^-}\in  \mathbb{R}^{N\times M}$ proportional to electrons for each $(i,j)$-th pixel as follows:
\label{review:19R1}\begin{equation}\label{eq:photonstoelectrons}
 I_{e^-}  = I_{ph} \cdot QE ,
\end{equation}
where $QE$ is the quantum efficiency [e$^-$/incident photons] for the given wavelength. The quantum efficiency for a photosensor is a function of wavelength and is usually provided by the manufacturer.

\subsubsection{Simulation of photo response non-uniformity}\label{subsec:prnusimulation}
The photo response non-uniformity (PRNU) is considered in our numerical model as a temporally-fixed light signal non-uniformity. According to our experimental results (see Subsection~\ref{subsec:prnudiscussion}), the PRNU can be modelled using a Gaussian distribution for each $(i,j)$-th pixel of the matrix $I_{e^-}$:\label{review:11R1}\label{review:12-1R1}
\begin{equation}
I_{PRNU.e^-}  = I_{e^-} +I_{e^-} \cdot \mathcal{N}(0,\sigma_{PRNU}^2)
\end{equation}
\label{review:6R1}where $\sigma_{PRNU}$ is the PRNU factor value. Since the PRNU does not change from one frame to the next, the matrix $\mathcal{N}$ is calculated once and then saved to a file. \label{review:21-4R1} In the simulations, the matrix  $\mathcal{N}$ can be loaded again to calculate the photo response non-uniformity. Therefore, since the same matrix is used for all simulations of the photo response non-uniformity, it will always remain the same as it should be.

\subsubsection{Simulation of dark signal}\label{subsec:darkcurrentsignalsim}
The dark   signal is calculated using Eq.~\ref{eq:darkcurrentrate} for all pixels in the  model. It is implemented using \texttt{ones} function in MATLAB as a matrix of the same size as the simulated photosensor. For each $(i,j)$-th pixel we have:
\begin{equation}\label{eq:darkcurrentsignalsimulation}
 I_{dc.e^-} = t_I\cdot D_R, 
\end{equation}
where $D_R$ is the average dark current from Eq.~\ref{eq:darkcurrentrate}, repeated here for convenience:
$$ D_R = P_A D_{FM} T^{3/2} \exp\left(-\frac{E_{gap}}{2\cdot k T}\right).$$

In this article we conduct all measurements at the room temperature $+25^{\circ}$ C. The matrix $I_{dc.e^-}$ represents a constant value of dark signal generated during the integration time $t_I$. However, since the electrons are generated randomly, the dark   signal is a subject of a \textit{dark current shot noise}.

\subsubsection{Simulation of dark current shot noise}\label{subsec:darkcurrentshotnoisesimulation}
\label{review:22-1R1}Similar to the photon shot noise, the \textit{dark current shot noise} is due to the random arrival of the generated electrons and therefore described by a Poisson process. The dark current shot noise can be simulated as a Poisson process $\mathcal{P}$   with mean $\Lambda$ as follows:\label{review:11-1R1}
\begin{equation}\label{eq:dshotnoisesimulation}
 I_{dc.shot.e^-}  = \mathcal{P}(\Lambda), \,\,\,\,\mbox{ where   } \Lambda = I_{dc.e^-}.
\end{equation}
We use the \texttt{poissrnd} function that generates Poisson random numbers with mean $\Lambda$ for each $(i,j)$-th element of the matrix $I_{dc.e^-}$ from Eq.~\ref{eq:darkcurrentsignalsimulation} that contains dark signal in electrons. The input of the \texttt{poissrnd} function will be the matrix $I_{dc.e^-}$ that contains the number of electrons, and the output will be the matrix $I_{dc.shot.e^-}$ with added dark current shot noise. \label{review:21-6R1} The matrix $I_{dc.shot.e^-}$ that corresponds to the dark current shot noise is recalculated each time the simulations are started, which corresponds to the temporal nature of the dark current shot noise.

\subsubsection{Simulation of dark current fixed pattern noise}\label{subsec:darkcurrentFPNsim}
Following the discussion in Subsection~\ref{subsec:DFPN}, the dark current Fixed Pattern Noise (FPN) is simulated using non-symmetric distributions to account for the ``outliers'' or ``hot pixels''. It is usually assumed that the dark current FPN can be described by Gaussian distribution. However, such an assumption provides a poor approximation of a complicated noise picture. This is shown in the experimental results in Subsection~\ref{subsec:darkfpnperf}.

\label{review:31-6R1}Studies show~\cite{comprehcmosmodel,baer2006model} that a more adequate model of dark current FPN is to use non-symmetric probability distributions. The concept is to use two distributions to describe very ``leaky'' pixels that exhibit higher noise level than others. The first distribution is used for the main body of the dark current FPN, with a uniform distribution  superimposed to model ``leaky'' pixels. For  simulations at room-temperature ($25^\circ$ C) authors in~\cite{comprehcmosmodel} use a \textit{logistic distribution}, where a higher proportion of the population is distributed in the tails~\cite{balakrishnanlogisticshandbook}. For higher temperatures, inverse Gaussian~\cite{chhikara1989inverse} and Log-Normal~\cite{aitchison1957lognormal} distributions have been proposed. It was reported~\cite{baer2006model} that the Log-Normal distribution works well for conventional 3T APS CMOS sensors with comparatively high dark current.

In our simulations we use the Log-Normal distribution for the simulation of dark current FPN in the case of short integration times, and superimposing other distributions for long integration times. The actual simulation code implements various models, including Log-Normal, Gaussian, and Wald distribution to elumate the dark current FPN noise for short- and long-term integration times.

The dark current FPN for each pixel of the matrix $I_{dc.shot.e^-}$ from Eq.~\ref{eq:dshotnoisesimulation} is computed as follows\label{review:31-5R1}:
\begin{equation}\label{eq:darkfpnsimulation}
I_{dc.FPN.e^-}  = I_{dc.shot.e^-}  + I_{dc.shot.e^-} \cdot ln\mathcal{N}(0,\sigma_{dc.FPN.e^-}^2)
\end{equation}
where $\sigma_{dc.FPN.e^-} = t_I D_R  D_N$ is from Eq.~\ref{eq:darkcurrentFPN}, $D_R$ is the average dark current, and $D_N$ is the dark current FPN factor. Since the dark current FPN does not change from one frame to the next,  the matrix $ln \mathcal{N}$ is calculated once and then can be re-used similar to the PRNU simulations in Subsection~\ref{subsec:prnusimulation}. 

The experimental results provided in Subsection~\ref{subsec:darkfpnperf} confirm that non-symmetric models, and in particular the Log-Normal distribution,  adequately  describe the dark current FPN in CMOS sensors, especially in the case of a long integration time (longer than 30-60 seconds). Our results are consistent with~\cite{baer2006model}; however, as we show in Subsection~\ref{subsec:verylongfpn}, using only one distribution cannot adequately describe the dark current FPN statistics. For long-exposure case, one needs to superimpose two (or more, depending on the sensor) probability distributions, as will be shown in Subsection~\ref{subsec:darkfpnperf}.

\subsubsection{Simulation of Source follower noise}\label{sec:sf:subsec:sfnoisesimulate}
The components of the source follower noise were discussed in Subsection~\ref{sec:SFnoise}. In the simulations, the source follower noise can be approximated as:
\begin{equation}\label{eq:sigma_sf}
\sigma_{SF} \approx \frac{\sqrt{
\sum\limits_{f=1}^{f_{clock}} S_{SF}(f)\cdot H_{CDS}(f) }}{A_{SN}A_{SF}(1-\exp[-t_s/\tau_D])},
\end{equation}
where $\sigma_{SF} $ is the source follower noise [$e^-$ rms], $f_{clock}$ is the readout clock frequency (typically several MHz), $t_s$ is the CDS sample-to-sampling time [sec], $A_{SN}$ is a sense node conversion gain from Eq.~\ref{eq:conversiontovoltage}, $A_{SF}$ is a source follower gain from Eq.~\ref{eq:conversiontosourcefollower}, $\tau_D$ is the Correlated Double Sampling (CDS) dominant time constant usually~\cite{janesickscintificCCD}  $\tau_D = 0.5t_s$ [sec]. In order to approximate the source follower noise, we need to calculate the power spectrum of the noise $S_{SF}(f)$ and the CDS transfer function $H_{CDS}(f)$.

The power spectrum $S_{SF}(f)$ of the source follower noise  can be approximated~\cite{photontransferbook} as:
\begin{equation}\label{eq:s_sf}
 S_{SF}(f) = W(f)^2 \cdot \left(1 + \frac{f_c}{f}\right)+S_{RTN}(f),
\end{equation}
where $W(f)$ is the thermal white noise in $V/Hz^{1/2}$ (typically $15 nV/Hz^{1/2}$), $f_c$ is the flicker noise corner frequency in [Hz], and $S_{RTN}(f)$ is the random telegraph noise (RTN) power spectrum that is given by~\cite{photontransferbook}:
\begin{equation}
 S_{RTN}(f) = \frac{2\Delta I^2 \tau_{RTN}}{4+(2\pi f \tau_{RTN})^2},
\end{equation}
where $\tau_{RTN}$ is the Random Telegraph Noise (RTN) characteristic time constant [sec] and $\Delta I$ [A] is the source follower current modulation induced by RTN. These constants are usually provided in the photosensor specifications and therefore can be used for the simulations.

The $H_{CDS}(f)$ is the CDS transfer function and is given by the following expression~\cite{photontransferbook}:
\begin{equation}\label{eq:h_cds}
 H_{CDS}(f) = \left[ \frac{1}{  1+(2\pi f \tau_D)^2  }\right] \cdot [ 2-2\cos(2\pi f t_s) ].
\end{equation}

In CCD photosensors, source follower noise is typically limited by the flicker noise. On the other hand, in CMOS photosensors the source follower noise is typically limited by the RTN noise.

\label{review:4-1R1}
Using parameters provided in the specifications for the photosensors, one has to calculate   Eq.~\ref{eq:sigma_sf} using the vector  $H_{CDS}(f)$ from Eq.~\ref{eq:h_cds} and $S_{SF}(f)$ from Eq.~\ref{eq:s_sf}. The functions $S_{SF}(f)$,  $S_{RTN}(f)$ , and $H_{CDS}(f)$ can be calculated   for each frequency value $f=1\dots f_{clock}$ using vector notation to streamline the simulation code. We assume the source follower noise to be additive and Gaussian-distributed:
\begin{equation}\label{eq:sourcefollowernoisesimulation}
 I_{SF.e^-}  = round[ \mathcal{N}(0,\sigma_{SF}^2)],
\end{equation}
where the standard deviation of source follower noise $\sigma_{SF}$ is given by Eq.~\ref{eq:sigma_sf}. The computed matrix $I_{SF.e^-}$ represents the source follower noise (in electrons).

\subsection{Model: From Charge to Voltage}\label{sec:chargevoltagesimulation}
\label{review:14R1}The simulation of charge to voltage conversion is performed as follows. The light signal $I_{light.e^-}$ contains photon shot noise (already added to the matrix $I_{e^-}$ from Eq.~\ref{eq:photonstoelectrons}) and the PRNU:\label{review:11-2R1}

\begin{equation}
I_{light.e^-} = I_{e^-}\cdot(1+ \mathcal{N}(0,\sigma_{PRNU}^2))
\end{equation}

The dark signal matrix $I_{dark.e^-}$ consists of dark signal, dark current shot noise, dark current FPN from Eq.~\ref{eq:darkfpnsimulation},  and source follower noise $I_{SF.e^-}$ is from Eq.\ref{eq:sourcefollowernoisesimulation} as follows:

\begin{equation}
I_{dark.e^-} =  I_{dc.shot.e^-}(1+ ln\mathcal{N}(0,\sigma_{dc.FPN.e^-}^2)) + I_{SF.e^-},
\end{equation}

Then the matrices that correspond to the light signal $I_{light.e^-}$ and dark signal $I_{d.e^-}$ are   summed   together and rounded:

\begin{equation}\label{eq:totalsignalcollected}
I_{total.e^-} = round \left( I_{light.e^-} + I_{dark.e^-} \right).
\end{equation}

\label{review2JEI:R1_Q1}The number of electrons is then truncated to the full well (the maximum number of electrons in the pixel) and rounded. Then the V/$e^-$ non-linearity is applied (if desired) to this sum using Eq.~\ref{eq:sensenodenonlin} and converted to voltage by multiplication by the sense node conversion gain $A_{SN}$. The details are provided in the following subsections.

\subsubsection{Simulation of sense node reset noise (kTC noise)}\label{subsec:simulationSNresetnoise}
The kTC noise is occurs in CMOS sensors (as we mentioned before in Subsection~\ref{sec:subsection:snresetnoise}), while for CCD sensors the sense node reset noise is removed~\cite{photontransferbook} by Correlated Double Sampling (CDS). Random fluctuations of charge on the sense node during the reset stage result in a corresponding photodiode reset voltage fluctuation. The sense node reset noise is given by   Eq.\ref{eq:ktcnoise}, repeated here for convenience:

$$\sigma_{RESET}  = \sqrt{\frac{k_B T}{C_{SN}}}.$$

The simulation of the sense node reset noise may be performed as an addition of non-symmetric probability distribution to the reference voltage $V_{REF}$. However, the form of distribution depends on the sensor's architecture and the reset technique~\cite{TurchettaCMOS}. An Inverse-Gaussian distribution can be used for the simulation of kTC noise which corresponds to a hard reset technique in the CMOS sensor, and the Log-Normal distribution can be used for soft-reset technique. The sense node reset noise can be simulated for each $(i,j)$-th pixel for the soft-reset case as:
\begin{equation}\label{eq:snresetsimulation}
 I_{SN.reset.V}  = ln\mathcal{N}(0,\sigma_{RESET}^2),
\end{equation}
where $\sigma_{RESET}$ is from Eq.\ref{eq:ktcnoise} and then added to the matrix $I_{REF.V}$ in Volts that corresponds to the reference voltage.

\subsubsection{Simulation of offset fixed pattern noise}\label{sec:subses:elgamalfpn}
A model of the offset FPN in CMOS sensors was represented~\cite{elgamalFPNmodeling} as the sum of two components: a column and a pixel component. Each component is modelled by a first order isotropic autoregressive process, where the processes are assumed uncorrelated. The model~\cite{elgamalFPNmodeling} uses autoregressive processes to model the offset FPN since the parameters can be easily and efficiently estimated from the hardware sensor data~\cite{elgamaluwideDRsensor}.\label{review:36R1}

\label{review:31-7R1}The column offset FPN from the model~\cite{elgamalFPNmodeling} can be described as follows. The noise is generated for each $j$-th column index of the matrix $I_V$ that corresponds to the voltage signal of the photosensor. The model assumes that the column offset FPN is a first order isotropic autoregressive process of the form:
\begin{equation}\label{eq:offsetFPNnoisesimulation}
I_{offset.FPN.V}(j) = a\bigl(I_V(j-1) + I_V(j+1)\bigr) + U(j),
\end{equation}
where the $U(j)$ are zero mean, uncorrelated random variables with the variance $\sigma_U$ , and $a\in [0,0.5]$ is a parameter that characterises the dependency of $I_V(j)$ on its two neighbours~\cite{elgamalFPNmodeling}. The data in the matrix $I_V(j)$ the repeated for every row to make the matrix of the appropriate size. More details are provided in~\cite{elgamalFPNmodeling}.

\subsubsection{Conversion to Voltage}\label{sec:simulationSourceFollower}
The simulation of the charge-to-voltage conversion uses the sense node gain $A_{SN}$ [V/$e^-$] as a parameter that is in the range $[1\dots 5] \mu V/e^-$ (sensor-dependent). Conversion from charge to voltage is performed in the sense node as follows:
\begin{equation}\label{eq:conversiontovoltage}
I_{SN.V} = V_{REF} - I_{total.e^-}\cdot A_{SN}
\end{equation}
where $I_{SN.V}$ is the matrix of sense node voltages, $V_{REF}$ is the reference voltage, $I_{total.e^-}$ is the matrix that corresponds to the total number of electrons collected by the ``software'' sensor (see Eq.\ref{eq:totalsignalcollected}) during the integration time, and $A_{SN}$ is the sense node gain in [V/$e^{-}$].

In case when the sense node reset noise (kTC noise) is simulated, the matrix that contains the reset noise $I_{SN.reset.V}$ from Eq.~\ref{eq:snresetsimulation} is added to the $V_{REF}$ in Eq.~\ref{eq:conversiontovoltage}:
\begin{equation}
 V_{REF} =  V_{REF}  + I_{SN.reset.V}.
\end{equation}

After that, the source follower gain $A_{SF}$ [V/V] is applied:
\begin{equation}\label{eq:conversiontosourcefollower}
I_{SF.V} = I_{SN.V}\cdot A_{SF}
\end{equation}
where $A_{SF}$ is a source follower gain [V/V]. 

If the offset FPN is simulated, the matrix $I_{offset.FPN.V}$ that corresponds to the offset FPN noise is added as follows:
\begin{equation}\label{eq:sourcefollowervoltage}
I_{SF.V} = I_{SF.V} + I_{offset.FPN.V}.
\end{equation}
Next, the V/V non-linearity from Eq.~\ref{eq:vvnonlin} is applied if necessary and the resulting matrix of voltages is stored in $I_{V}$. After that, the matrix $I_{V}$ that corresponds to voltages is converted to digital numbers.

\subsubsection{Simulation of the V/$e^-$ gain non-linearity}\label{review:11-3R1}
For the simulation model, the V/$e^-$ non-linearity can be expressed from   Eq.\ref{eq:venonlinsignal} as:
\begin{equation}\label{eq:sensenodenonlin}
 I_{SN.V} = V_{REF}\exp\Bigl[ -\frac{\alpha\cdot I_{total.e^-}\cdot q }{k_1 } \Bigr]
\end{equation}
where $q$ is the charge of an electron, and $\alpha$ is the coefficient of non-linearity. The parameter $\alpha$ is usually provided in the photosensor specifications.

\subsubsection{Simulation of the V/V gain non-linearity}\label{subses:vvnonlinsimulation}
The voltage matrix is multiplied by the new source follower gain $A_{SF_{new}}$:
\begin{equation}\label{eq:vvnonlin}
 I_{SF.V} = I_{SN.V}\cdot A_{SF_{new}}
\end{equation}
where $I_{SN.V}$ is the sense node voltage signal,  $I_{SF.V}$ is the source follower voltage signal, and $A_{SF_{new}}$ is a new source follower gain according to Eq.\ref{eq:sfnewgain} repeated here for convenience:
$$A_{SF_{new}} = (\gamma_{nlr} -1) \cdot \frac{V_{SN}}{V_{FW} } + A_{SF}.$$
The next step involves the ADC for quantisation to a digital signal.

\subsubsection{Simulation of correlated double sampling}\label{subsec:CDSsimulation}
\label{review:3-1R1}In this paper, we use the algorithm of correlated double sampling (CDS) that is similar to the method described in~\cite{farrell2004simulation}. The algorithm is as follows. During the   simulations, two images are acquired: a light image that contains both signal and noise, and a dark image with zero exposure time. The CDS algorithm then subtracts these two images to eliminate fixed pattern noise.

In the real hardware sensor, the CDS techniques can be different and more sophisticated. This can contribute to mismatch between the data from the real sensor and from the numerical simulator. That is, the CDS algorithms in the simulations can be less aggressive and therefore the residual noise on the image can be larger than on the actual sensor.

\subsection{Model: From Voltage to Digital Numbers}\label{sec:fromvoltagetodn}
After the electron matrix has been converted to voltages, the sense node reset noise and offset FPN noise are  added, the V/V gain non-linearity is applied (if desired), the ADC non-linearity is applied (if necessary). Finally the result is multiplied by ADC gain and rounded to produce the signal as a digital number:
\begin{equation}\label{eq:adcconvertersimulation}
I_{DN} =  round (A_{ADC}\cdot I_{total.V}),
\end{equation}
where $I_{total.V} = (V_{ADC.REF} - I_{V})$ is the total voltage signal accumulated during one frame acquisition, $V_{ADC.REF}$ is the maximum voltage that can be quantified by an ADC, and $I_V$ is the total voltage signal accumulated by the end of the exposure (integration) time and conversion. Usually $I_V = I_{SN.V}$ from Eq.~\ref{eq:sourcefollowervoltage} after the optional V/V non-linearity is applied. In this case, the conversion from voltages to digital signal is linear. The non-linear case is considered below.

\subsubsection{ADC non-linearity simulation}\label{subsec:adcnonlinandnoise}
In our model, we simulate the Integral Linearity Error (ILE) of the ADC as a dependency of ADC gain $A_{ADC.linear}$ on the signal value. Denote $\gamma_{ADC.nonlin}$ as an ADC non-linearity ratio (e.g., $\gamma_{ADC.nonlin} = 1.04$). The linear ADC gain can be calculated from Eq.~\ref{eq:kadc} as $A_{ADC} = 1/K_{ADC}$ and used as $A_{ADC.linear}$. The non-linearity coefficient $\alpha_{ADC}$ is calculated as:
\begin{eqnarray}\label{eq:kadcnonlincalc}
\alpha_{ADC} =  \left( \frac{ \log(\gamma_{ADC.nonlin} \cdot A_{ADC.linear} )}{\log(A_{ADC.linear})} - 1 \right)* \\ \nonumber
*\frac{1}{V_{ADC.REF}}
\end{eqnarray}
where $V_\mathrm{ADC.REF}$ is the maximum voltage that can be quantified by an ADC: 
\begin{eqnarray}\label{eq:kadcnonlin}
A_{ADC.nonlin} = A_{ADC.linear}^{1-\alpha_{ADC} I_{total.V}},
\end{eqnarray}
where $A_{ADC.linear}$ is the linear ADC gain. The new non-linear ADC conversion gain $A_{ADC.nonlin}$ is then used for the simulations for Eq.~\ref{eq:adcconvertersimulation}.

\subsubsection{Simulation of quantisation noise}\label{subsec:adcnoise}
It is assumed that the quantisation error is uniformly distributed between -0.5 and +0.5 of the LSB and uncorrelated with the signal.  Denote $q_{ADC}$ the quantising step of the ADC. For the ideal DC, the quantisation noise is:
\begin{equation}
 \sigma_{ADC} = \sqrt{ \frac{q_{ADC}^2 }{12}}.
\end{equation}
If $q_{ADC} = 1$ then the quantisation noise is $\sigma_{ADC} = 0.29$ DN. The quantisation error has a uniform distribution. We do not assume any particular architecture of the ADC in our high-level sensor model.

\section{Experimental validation of the photosensor model}\label{sec:experimentalvalidation}
Using the formulated   photosensor model, we were able to simulate realistically noised images similar to those obtained from a hardware CMOS photosensor. A comparison with custom-made 5T CMOS photosensor with $1300\times 1900$ pixels of size $5.7\mu m$ is provided to validate the developed numerical model.

\label{review:10R2}The simulations were performed as follows. The parameters of the hardware CMOS sensor were taken from the specifications (see Table~\ref{tab:photosensorsim}) provided by the manufacturer and used in the simulations. A series of the measurements were conducted both for the hardware sensor and the simulated ``software'' photosensor.

The methods and procedures of the measurement were the same for the hardware and the simulated photosensor. For example, in case of estimation of the radiometric function, the flat field for the hardware sensor was generated by an array of LEDs. The measurements were performed for monochromatic light with wavelength $\lambda=0.55 \mu$ m (\label{review2JEI:R2Q9-2} we assume for the same of simplicty that the light is monochromatic, and therefore we need the spectral response of the sensor at one wavelength $\lambda=0.55 \mu$ m). \label{review:5-2R1} For the numerical model, the uniform image was used that corresponds to the light signal of the LEDs. \label{review2JEI:R2Q7} The measurements were conducted at room temperature ($+25^\circ$C) if not stated otherwise.

\label{text:nonlinearities}
\begin{table}[ht!]
\centering
\caption{Parameters of the  simulated CMOS sensor (taken from the manufacturer's specifications).}
	\label{tab:photosensorsim}
	\begin{tabular}{ll}
	\hline
	\hline
	\textbf{Sensor's }  & \textbf{Value} \\
	\textbf{parameter}  & \textbf{} \\
	\hline
	\hline
	number of pixels  & $1300\times 1900$\\
	\hline
	wavelength $\lambda$ & $0.55 \mu m$\\
	\hline
	pixel size & $5.7 \mu m$ \\
	\hline
	pixel fill factor& 50\% \\
	\hline
	Full well & 33000 $e^-$ \\
	\hline
	QE &  0.60 \\
	\hline
	PRNU factor & 0.5\% \\
	\hline
	dark current FPN factor & 1\% \\
	\hline
	Column offset FPN factor& 0.10\% \\
	\hline
	dark current figure of merit &  1.00 nA/$cm^2$ \\
	\hline
	Sense node gain &  5.00 $\mu V/e^-$\\
	\hline
	Clock speed	& $20$ MHz \\
	\hline
	ADC bit &  12 bit\\
	\hline
	\hline
	\end{tabular}
\end{table}

\label{review:10-1R2}Not all the parameters of the hardware photosensor were provided by the manufacturer. For example, the ADC non-linearity\label{review:9R1} (specifically, Integral Linearity Error) was not provided; later, the manufacturer stated that a low-grade ADC with $ILE \sim 4\%$ has been used. Also it was reported by the manufacturer that the source follower may have an uncompensated non-linearity as well. Therefore, we provide the results of the simulation for the linear model (all the non-linearity turned off) and the non-linear model (ADC and V/V non-linearity is turned on).

\label{review2JEI:R2Q8}The goal is to provide an illustration that even in the case of incomplete information from the manufacturer's specifications, the model parameters can be adjusted for the consistency with the hardware sensor. This allows obtaining realistically noised images with the statistical properties that are close to the hardware photosensor.

Our simulations of the CMOS sensor include the following sources of noise: photon shot noise, photo response non-uniformity, dark   signal, dark current shot noise, dark current Fixed Pattern Noise, source follower noise, offset FPN, ADC quantisation noise. \label{review:16R2}The ADC non-linearity and V/V non-linearity were simulated.

\subsection{Radiometric function}\label{subsec:radiometricfunc}
The radiometric function of the hardware photosensor was estimated and compared with the simulated sensor. In real experiments, we used an array of the green LEDs to form a flat-field scene. The light of green LEDs is passed through a ground glass to eliminate flat-field non-uniformity. \label{review2JEI:R2Q10} The camera in these experiments did not have lens. In the simulations we use uniform images that correspond to a flat-field light scene.

\paragraph*{Method of measurements:}
The images were taken over a range of integration times to cover the whole dynamic range of a photosensor (both hardware and simulated sensors). An area of $512 \times 512$ pixels from the centre of each acquired image was used for the analysis. The mean and the standard deviation values of that $512 \times 512$ pixel area were calculated for each integration time. \label{review:11R2}The mean value was then used as a signal value for the radiometric function and the standard deviation was considered as a result of the photosensor noise.

\begin{figure}[ht!]
\centering \includegraphics[width=0.99\linewidth]{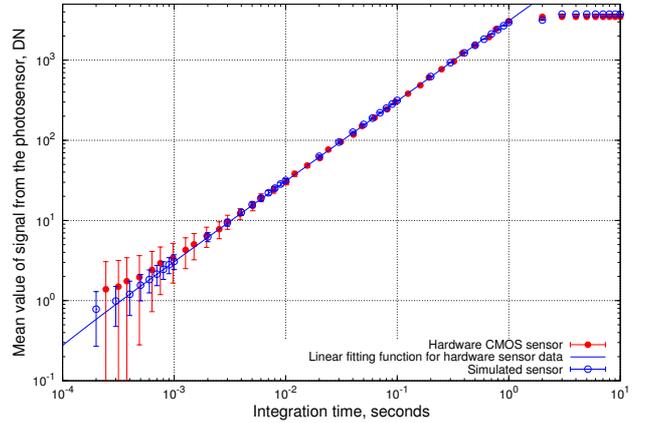}
	\caption{The radiometric function for the simulated photosensor and the hardware CMOS sensor.}
	\label{fig:plotRadiometryFunciton}
\end{figure}

\paragraph*{Results:}\label{review:12R2}
The results are presented in Fig.~\ref{fig:plotRadiometryFunciton}, where the simulated sensor data are marked by a ``$\circ$'', and the data from the hardware CMOS sensor are marked by a ``$\bullet$''.

Although the radiometric function for the simulated sensor exhibits a similar behaviour compared with the hardware sensor, one can note from the Fig.~\ref{fig:plotRadiometryFunciton} that the standard deviation is higher for the hardware sensor compared to the simulated sensor at low integration times. Such differences can be explained by the fact that we rely on the manufacturer's specifications for the sensor parameters. 

\label{review:13-1R2}The fact that the noise of the hardware sensor is larger than the one for the model can be also attributed to the CDS (correlated double sampling) algorithm. The details of the CDS in the hardware sensor were not disclosed by the manufacturer, and the algorithm that we implemented (see Subsection~\ref{subsec:CDSsimulation}) is probably more aggressive than the actual algorithm in the hardware sensor. On the averaged dark frames, some offset FPN artefacts are still identifiable and definitely contributing to the noise picture. Nonetheless, the behaviour of the numerical model can be considered as consistent with the hardware sensor.

\subsection{Photon Transfer Curve}\label{subsec:ptcsim}
As above, the parameters of the hardware photosensor were taken from the manufacturer's specifications in Table~\ref{tab:photosensorsim} and substituted into the numerical simulation software. The method of Photon Transfer Curve (PTC) measurement described below was the same for both numerical model and the hardware sensor.

\paragraph*{Method of measurements:}
Usually the data for the Photon Transfer Curve (PTC) are obtained from the measurements of the radiometric function~\cite{janesickscintificCCD,liebe1998active}. We use different method of measurements for the PTC, which allows obtaining more data points for PTC. We made the input scene that covers the whole dynamic range of the photosensor and set the integration time as $t_I = 100\,$ msec. Such a scene can be visually described as a vertical or horizontal linear gradient, and therefore covers all the signal values from dark noise to sensor saturation. 

\begin{figure}[ht!]
\centering
\subfigure[]{
\includegraphics[width=0.99\linewidth]{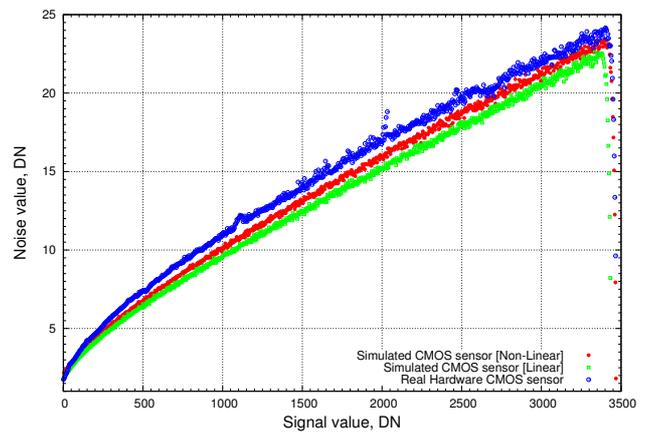}
\label{fig:plotPTClinearmultipleSignal-vs-noise-linear-3in1}
}
\subfigure[]{
\includegraphics[width=0.99\linewidth]{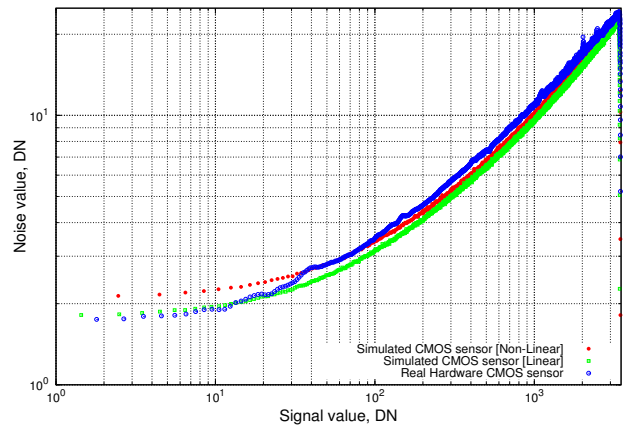}
\label{fig:plotPTCmultipleSignal-vs-noise-loglog-3in1}
}
\caption{Photon Transfer Curves of hardware CMOS sensor and the simulated sensor: \subref{fig:plotPTClinearmultipleSignal-vs-noise-linear-3in1} linear scale, and \subref{fig:plotPTCmultipleSignal-vs-noise-loglog-3in1} logarithmic scale. Hardware CMOS sensor data is marked by \textcolor{Blue}{``$\circ$''} symbol; results of the linear model are marked by \textcolor{Green}{``$\square$''}, and the non-linear model is marked by \textcolor{Red}{``$\bullet$''} (colour online). }
\label{fig:plotPTC_snr_vs_sigmultipleSignal-3in1}
\end{figure}

\label{review2JEI:R2Q12}We took $K=64$ images (matrices $S_{1}\dots S_{K}$ of size $N\times M$ pixels) of such a gradient input scene, and averaged the images to reduce temporal noise, producing the matrix $S_{mean}$ of size $N\times M$ pixels. Now, the averaged image $S_{mean}$ contains the whole dynamic range of the signal values produced by a sensor. In order to plot the PTC, we need to find the corresponding standard deviation values (i.e., noise) for each value of the mean signal in the image $S_{mean}$. Denote the matrix of standard deviations $S_{std}$. For each $(i,j)$-th pixel in the matrix $S_{mean}$,  we can find the standard deviation by selecting $(i,j)$-th pixels from all $K$ images $S_{1}\dots S_{K}$. That is, we create a $1\times K$ vector $S_{std}^{ij}$, and each $k$-th value of this vector corresponds to $(i,j)$-th pixel of a $k$-th image from the set $S_{1}(i,j)\dots S_{K}(i,j)$. Then we calculate the standard deviation $S_{std}(i,j) = std\left( S_{std}^{ij} \right)$. Thus for each  $(i,j)$-th signal value in $S_{mean}$, there is a corresponding  $(i,j)$-th value of standard deviation from the array $S_{std}$ (that is, a noise value). The data from these two matrices is plotted as the PTC (see Fig.~\ref{fig:plotPTC_snr_vs_sigmultipleSignal-3in1}).

% There were two arrays of data evaluated from the averaged images for each pixel. The first array contained mean values of pixels $S_{mean}$, and the second array contained standard deviations of pixels $S_{std}$. The fixed range of values in digital numbers (DN, i.e., digital signal) was chosen for the mean values of pixels (e.g., the values of the signal were taken in the range between 2~DN and 4~DN, then between 4~DN and 6~DN and so on). For each fixed range (1~DN in our case), the mean values of the signal were taken from the array $S_{mean}$. For each signal value in $S_{mean}$, the corresponding standard deviation value from the array $S_{std}$ was taken as an estimation of the sensor noise.

% This can be seen in Fig.~\ref{fig:plotPTCmultipleStatisticsPTC}: where there are at least 100 data points for each signal value in digital numbers (DN) that will be used for the PTC plotting. 
% 
% \begin{figure}[ht!]
% \centering\includegraphics[width=0.49\linewidth]{plotPTCmultipleStatisticsPTC}
% \caption{Number of data points for the Photon Transfer Curve.}
% \label{fig:plotPTCmultipleStatisticsPTC}
% \end{figure}

This method allows obtaining a statistically significant amount of data points for the PTC plot.
% , provided that the size of the sensor is big enoug. 
% Second, since all the images are taken at one fixed (and preferably short) exposure time, the impact of the dark noise (especially dark current FPN) is minimised. 
Increased number of the data points in the PTC, in turn, may provide more information about the noise in the photosensor. This can be seen in Fig.~\ref{fig:plotPTC_snr_vs_sigmultipleSignal-3in1}, where the irregularities and noise spikes can be seen at the signal values 1100 and 2000 DN.

\paragraph*{Results:}
\label{review:16-2R2}The results of comparison of the data from the hardware photosensor and simulations for both linear and non-linear models are discussed below. Using the ADC non-linearity (described in Sec.~\ref{subsec:adcnonlinandnoise}, Eq.~\ref{eq:kadcnonlin}) and the V/V non-linearity (described in Sec.~\ref{subses:vvnonlinsimulation}, Eq.~\ref{eq:vvnonlin}, where $A_{SF_{new}}$ is a new source follower gain according to Eq.\ref{eq:sfnewgain}), we were able bring the model of the sensor closer to the hardware photosensor.

In order to show the impact of the non-linearity, we simulated V/V non-linearity with ratio $A_{SF_{nl}} = 1.05$ and an ADC-nonlinearity of ratio $\gamma_{ADC.nonlin} = 1.04$. \label{review:14R2} The linear and non-linear models are compared with the data from the hardware sensor on the Photon Transfer Curve in Fig.~\ref{fig:plotPTC_snr_vs_sigmultipleSignal-3in1}. The data on the PTC is as follows:
\begin{enumerate}
 \item \textit{hardware data} ( ``$\circ$'' in Fig.~\ref{fig:plotPTC_snr_vs_sigmultipleSignal-3in1});
 \item \textit{fully linear model} ( ``$\square$'' in Fig.~\ref{fig:plotPTC_snr_vs_sigmultipleSignal-3in1}) non-linearity is turned off;
\item  \textit{non-linear model} ( filled ``$\bullet$''  in Fig.~\ref{fig:plotPTC_snr_vs_sigmultipleSignal-3in1}) -- the model uses both the V/V and the ADC non-linearity\label{review:16-4R2} (the V/V non-linearity is described in Sec.~\ref{subses:vvnonlinsimulation}, and the ADC non-linearity is described in Sec.~\ref{subsec:adcnonlinandnoise}). 
\end{enumerate}

Comparing the photon transfer curves on Fig.~\ref{fig:plotPTC_snr_vs_sigmultipleSignal-3in1}, one can see that the non-linearities model is closer to the hardware sensor. As a short summary, the results presented above show that the model of the sensor is consistent with the experimental data from the hardware sensor.

\subsection{Signal-To-Noise Ratio}\label{subsec:snrmeasurements}
\label{review:15-1R2}The signal-to-noise ratio (SNR) has been estimated for both model and the hardware sensor. The same method of measurements was used for the hardware sensor and the model, thus providing the means to compare the data.

\begin{figure}[ht!]
\centering
\subfigure[]{
\includegraphics[width=0.99\linewidth]{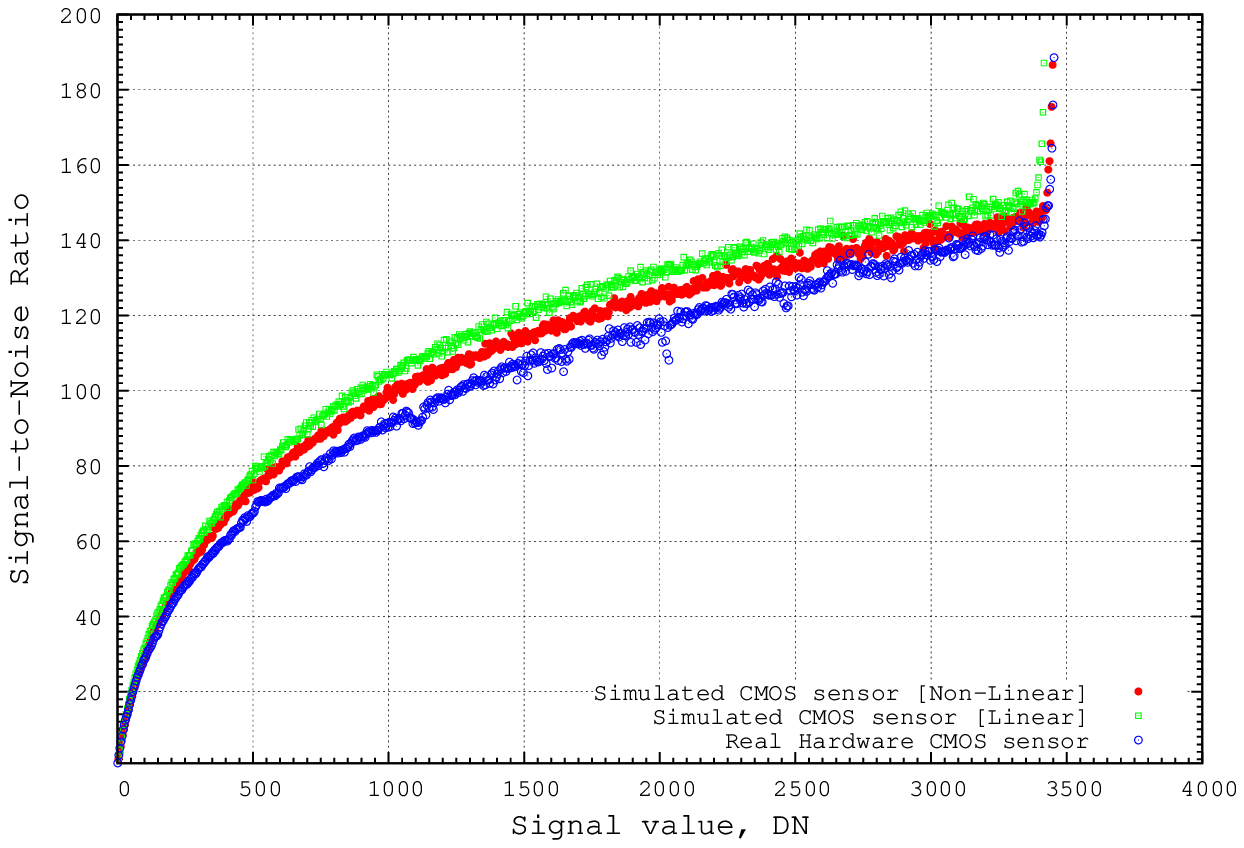}
\label{fig:plotPTC_snr_vs_sigmultipleSignal-vs-noise-linear-3in1}
}
\subfigure[]{
\includegraphics[width=0.99\linewidth]{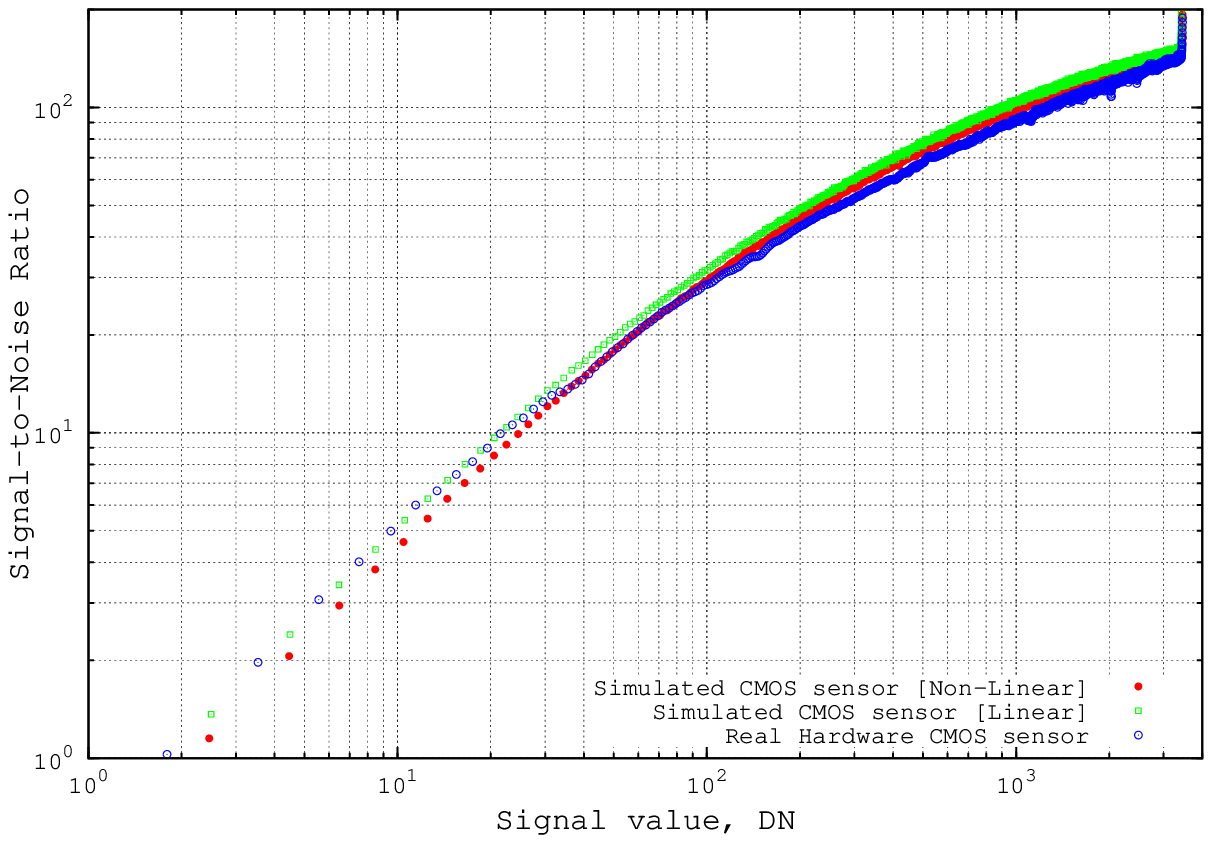}
\label{fig:plotPTC_snr_vs_sigmultipleSignal-vs-noise-loglog-3in1}
}
\caption{Dependence of signal-to-noise ratio versus signal for \subref{fig:plotPTC_snr_vs_sigmultipleSignal-vs-noise-linear-3in1}  linear  and \subref{fig:plotPTC_snr_vs_sigmultipleSignal-vs-noise-loglog-3in1} logarithmic scales.
Hardware CMOS sensor data are marked by \textcolor{Blue}{``$\circ$''}; results of fully linear model aremarked by \textcolor{Green}{``$\square$''}, and the non-linear model is marked by \textcolor{Red}{``$\bullet$''} (colour online). }
\label{fig:plotPTC_snr_vs_sigmultipleSignal-vs-noise-loglogandlinear-3in1}
\end{figure}

\paragraph*{Method of measurements:}
The method of measurements for the signal-to-noise ratio (SNR) estimation was the same as described in Subsection~\ref{subsec:ptcsim}. The data were derived from the PTC data and plotted as SNR versus signal values in Fig.~\ref{fig:plotPTC_snr_vs_sigmultipleSignal-vs-noise-loglogandlinear-3in1} for linear and logarithmic axis. Linear plot is more suited for the analysis of the SNR when the signal is large, while logarithmic plot gives more information for the a low signal.

Similarly to the PTC results, we use a qualitative comparison of PTC curves for the hardware and the model. The aim is to provide the means of model and hardware comparison in case when the only data available are the resulting images from the photosensor. However, it is difficult to distinguish between the different non-linearities from the resulting image only.

\paragraph*{Results:}
The comparison of SNR versus signal curves in Fig.~\ref{fig:plotPTC_snr_vs_sigmultipleSignal-vs-noise-loglogandlinear-3in1} gives a qualitative estimation of the influence of non-linearity on the Photon Transfer Curve. \label{review2JEI:R2Q14} The data and the fitting parameters are the same as in the previous section (presented in Fig.~\ref{fig:plotPTC_snr_vs_sigmultipleSignal-3in1} above).

One can compare from Fig.~\ref{fig:plotPTC_snr_vs_sigmultipleSignal-vs-noise-loglog-3in1} the hardware data  with two  simulated models:

\begin{enumerate}
 \item \textit{fully linear model} ( ``$\square$'' on Fig.~\ref{fig:plotPTC_snr_vs_sigmultipleSignal-vs-noise-loglog-3in1}) -- the  simulations in this case do not use the non-linearity: neither V/V nor ADC non-linearity was used.
\item  \textit{non-linear model} ( filled $\bullet$  on Fig.~\ref{fig:plotPTC_snr_vs_sigmultipleSignal-vs-noise-loglog-3in1}) -- the model uses both the V/V and the ADC non-linearity\label{review:16-1R2} (the V/V non-linearity is described in Sec.~\ref{subses:vvnonlinsimulation}, specifically Eq.~\ref{eq:vvnonlin}, where $A_{SF_{new}}$ is a new source follower gain according to Eq.\ref{eq:sfnewgain}; the ADC non-linearity is described in Sec.~\ref{subsec:adcnonlinandnoise}, specifically Eq.~\ref{eq:kadcnonlin}). 
\end{enumerate}

It can be seen that the fully linear model ( ``$\square$'') overestimates the signal-to-noise ratio of the sensor compared to the non-linear model ( filled $\bullet$). The non-linear model is closer to the hardware data and therefore better describes the photosensor. This is not surprising as the ADC of the camera and the source follower has an uncompensated non-linearity.

\label{review:7-comment2R1} From the comparison of the SNR curves in Fig.~\ref{fig:plotPTC_snr_vs_sigmultipleSignal-vs-noise-loglogandlinear-3in1} one can conclude that the non-linear model is closer to the hardware data. However, it is difficult to distinguish the influence of the non-linearities using only the image data. This is often the case when only the resulting images from the sensor are available for a researcher. This is the reason why we use the comparison of the PTC curves for model and hardware sensors: although such a method gives qualitative results, it is simple and informative enough to judge the consistency of the  model with the hardware.

\label{review:10R1}The difference between the hardware and simulated sensor's noise performance can be explained by the fact that the details of the CDS algorithm were not disclosed by the manufacturer.  Nonetheless, the non-linear model is closer with the hardware sensor and therefore leads to more realistically noised images that can be obtained by the numerical model.

\subsection{Photo Response Non-Uniformity}\label{subsec:PRNU}
\label{review:10-2R2}The method and the results of the PRNU measurements are reported in this subsection. The probability density and the PRNU factor were evaluated for both the numerical model and the hardware photosensor. \label{review2JEI:R2Q14-1} That is, we empirically measured the PRNU and showed that it is close to the manufacturers measurements.

\paragraph*{Method of measurements:}\label{subsec:prnumethod}
The procedure of the measurement of PRNU properties is as follows.  \label{review:8R1} The flat-field uniform scene was formed using an array of green LEDs for the hardware CMOS sensor. In order to reduce the illumination non-uniformity, a ground glass was used. In case of the model (``software photosensor''), the uniform images were generated and ``acquired'' in simulations. The exposure (integration) time was set to the value when the mean of acquired uniform images were equal to half of the saturation level of the sensor~\cite{prnuqualification}. The integration time was the same for the hardware photosensor and the simulations. 

Next, 64 images of the uniform flat field scene were acquired and averaged. The same number of the dark frames with the same exposure time were acquired and averaged. The averaged dark frame was subtracted from the averaged image of the uniform flat field image in order to eliminate the influence of the dark current FPN and the offset FPN on the PRNU measurements.

The mean, $\mu_{frame}$, and the standard deviation, $\sigma_{frame}$, were calculated from the averaged uniform flat field image. Standard deviation and mean values were estimated using maximum likelihood estimates (MLEs) for the parameters of a Gaussian distribution. The value of the PRNU factor was calculated as follows:
\begin{equation}\label{eq:prnuestimation}
PRNU = \frac{\sigma_{frame}} {\mu_{frame}} \%. 
\end{equation}
The PRNU factor for the numerical model was then compared with the specifications of the hardware sensor.

\paragraph*{Results:}\label{subsec:prnudiscussion}
\label{review:17R2}We estimated the PRNU value for the hardware CMOS photosensor from the mean value $\mu = 1520$ DN and the standard deviation $\sigma = 7.9$. These values were obtained maximum likelihood estimates (MLEs) for the parameters of a Gaussian distribution on 95\% confidence intervals. The PRNU factor for the hardware CMOS sensor was calculated according to Eq.~\ref{eq:prnuestimation} and found to be $PRNU_{hardware} \approx 0.52$\%. This result is consistent with the manufacturer specifications in Table~\ref{tab:photosensorsim}, where the $PRNU_{simulate} = 0.50$\% was stated. The same PRNU value was used for the simulations.

\begin{figure}[ht!]
\centering\includegraphics[width=0.99\linewidth]{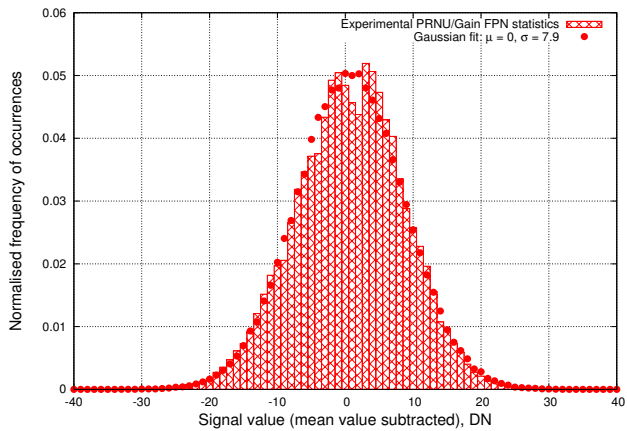}
\caption{Estimated probability density function of the Photo Response Non-Uniformity (mean value $\mu_{frame} = 1520$ is subtracted).}
\label{fig:plotPDFestimationmultiplesemilog}
\end{figure}

The PRNU statistical properties were also estimated for the hardware CMOS photosensor to confirm our assumption that the PRNU can be simulated using a Gaussian distribution. The results are presented in Fig.~\ref{fig:plotPDFestimationmultiplesemilog} where the mean value $\mu_{frame} = 1520$DN has been subtracted. As one can see in Fig.~\ref{fig:plotPDFestimationmultiplesemilog}, the PRNU can be assumed Gaussian with very good accuracy.

Note that the percentage of the PRNU in the signal should be constant until reaching the full well~\cite{janesickpersonal} value. In some articles (e.g. ~\cite{comprehcmosmodel}) it is stated that the percentage of PRNU depends on light irradiance. For signals less than approximately 10000 electrons, the photon shot noise is dominant, and the PRNU can be confused with the photon shot noise.

\subsection{Noise Spectrogram}\label{subsec:spectrogram}
Periodic variations of the noise can be characterised by computing a spectrogram, i.e., a power spectrum of the spatial variations, as described in the EMVA1288 Standard~\cite{EMVA1288}. The square root of the power spectrum is displayed as a function of the spatial frequency (in units of cycles per pixel) in the spectrogram. Typically the spectrogram is calculated for three conditions: complete darkness, when the signal is at 50\% of saturation, and when the signal is at 90\% saturation~\cite{EMVA1288}.

\paragraph*{Method of measurements for noise spectrogram}
The measurement approach to determine the noise spectrogram is based on the ``Spectrogram Method'' in the EMVA1288 standard~\cite{EMVA1288}. The spectrogram is computed by taking the mean of the amplitude of the FFT on each line of the image. First, the mean value of the image is computed $$\mu_y = \frac{1}{MN}\cdot \sum \limits_{n} \sum \limits_{m} y(n,m).$$ Then for the $j$-th line of the $M$ lines of the image, the amplitude of the FFT is computed.

The procedure is as follows. Make an array $y_j (k)$ of length $2N$, then copy pixels from the image to the first half of the array $(0 \leq k \leq N -1)$ and subtract the mean. Fill the second half of $y_j (k)$ with zeros $( N \leq k \leq 2N -1)$ and take FFT from $y(k)$ to obtain  $\underline{Y}_j (n)$.

Then the amplitude $S_j$ of the FFT 
$$ S(n) = \sqrt{ \frac{1}{M} \cdot \sum  S_j(n)^2 \,\,}\,\,\,$$
is computed for the $j$-th line (i.e., $S_j(n)$ is the amplitude of the Fourier transform of the $j$-th line).  The $N+1$ values $S(n)$ with $0 \leq n \leq N$ form the spectrogram of the image as a dependency of mean FFT amplitude versus spatial frequency. More details can be found in Subsection 7.3.2 ``Spectrogram Method'' in the EMVA1288 standard Release A2.01~\cite{EMVA1288}.

\begin{figure}[ht!]
\centering
\subfigure[]{
\includegraphics[width=0.99\linewidth]{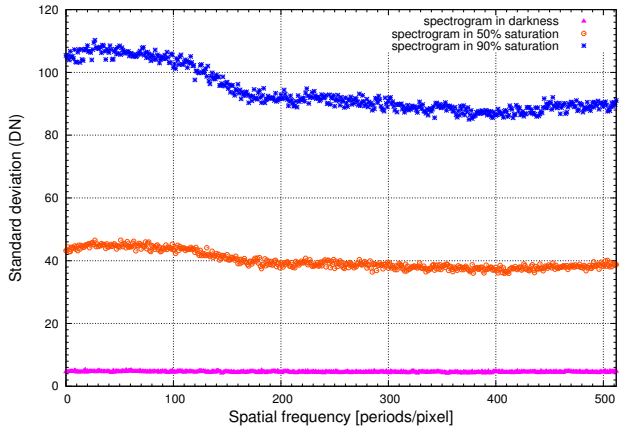}
\label{fig:plot3in13in1EMVA1288Spectrogrammhardware}
}
\subfigure[]{
\includegraphics[width=0.99\linewidth]{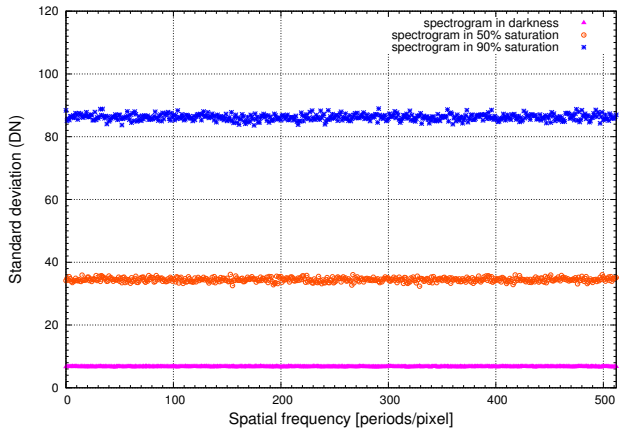}
\label{fig:plot3in13in1EMVA1288Spectrogrammsim}
}
\caption{Spectrogram of the noise taken in different light conditions (in darkness, at 50\% light saturation, and at 90\% light saturation) for:  \subref{fig:plot3in13in1EMVA1288Spectrogrammhardware} the hardware CMOS sensor; and \subref{fig:plot3in13in1EMVA1288Spectrogrammsim} the simulated photosensor.}
\label{fig:plot3in13in1EMVA1288Spectrogrammsim-3in1}
\end{figure}

\paragraph*{Results for Noise Spectrogram}
Three spectrograms were calculated for the simulated and the hardware CMOS photosensor: in darkness, when the integration time is set such that the sensor is at 50\% saturation with $t_{I_1} = 0.05 \, sec$, and when the integration time is set to produce 90\% saturation with $t_{I_2} = 0.1 sec$. The resulting spectrograms were merged into one plot for both the hardware and the simulated CMOS sensor and presented in Fig.~\ref{fig:plot3in13in1EMVA1288Spectrogrammhardware} and Fig.~\ref{fig:plot3in13in1EMVA1288Spectrogrammsim}, respectively.

According to the model~\cite{EMVA1288}, the spectrograms should be flat with occasional peaks only. The results of noise spectrogram (see Fig.~\ref{fig:plot3in13in1EMVA1288Spectrogrammsim-3in1}) are in agreement with the model~\cite{EMVA1288}. However, on the spectrograms from the real sensors (see Fig.~\ref{fig:plot3in13in1EMVA1288Spectrogrammhardware}) one can see non-flatness of the spectrogram for the 50\% saturation and 90\% saturation cases.

\paragraph*{Method of measurement for non-whiteness coefficient of noise}
If there is no spatial correlation in the noise (i.e., it is purely random) then the power spectrum should be flat (white spectrum). However, in a hardware photosensors spatial noise can be dominated by periodic artefacts, such as vertical or horizontal stripes in the image, that can be observed in the power spectrum and spectrogram (as seen in Fig.~\ref{fig:plot3in13in1EMVA1288Spectrogrammhardware}). To assess such non-whiteness quantitatively, a non-whiteness coefficient~\cite{EMVA1288} is usually calculated:
\begin{equation}\label{eq:nonwhitenesscoef}
 F = \frac{\sigma_{y.total}^2 }{ \sigma_{y.white}^2}
\end{equation}
and  $F\approx 1$ for white noise. The total noise variance is:
$$\sigma_{y.total}^2 = \frac{1}{N+1}\cdot  \sum \limits_{n} S(n)^2$$ is the variance of the noise including all artefacts. The $\sigma_{y.white}^2$  is the square of the height of the flat part seen in the spectrogram curve, it describes the white part of the photosensor's noise and can be estimated by taking the median of the spectrogram.

\paragraph*{Results for non-whiteness coefficient of noise}
The non-whiteness coefficient $F$ was calculated according to Eq.~\ref{eq:nonwhitenesscoef} for each spectrogram for both the hardware and simulated CMOS sensor. Uncertainty of the non-whiteness coefficient $F$ was estimated from the standard deviation of 100 measurements of $F$. The results of estimating the non-whiteness coefficient are summarised in Table~\ref{tab:spectrogramcalc}.

\begin{table}[ht!]
	\caption{Results of spectrogram estimation for the hardware sensor and the simulated photosensor.}
	\label{tab:spectrogramcalc}
	\begin{center}
	\begin{tabular}{rccc}
\hline\hline
Experi-	& int.	&\multicolumn{2}{c}{Non-whiteness coef.}\\
ment	& time, 	&\multicolumn{2}{c}{$F = \sigma_{y.total}^2 / \sigma_{y.white}^2$ for sensor:}\\
\cline{3-4}
type	& sec			& hardware & simulated \\
\hline\hline
darkness: 	& $0.05$	& $0.999  \pm 0.018 $	& $0.999  \pm 0.002 $	\\
darkness: 	& $0.1$	& $1.032  \pm 0.020 $	& $0.999  \pm 0.002 $ \\
\hline
50\% sat.:& $0.05$	& $1.062  \pm 0.018 $	& $0.999  \pm 0.002 $ \\
\hline
90\% sat.:& $0.1$	& $1.079  \pm 0.020 $	& $0.998  \pm 0.002 $\\
\hline\hline
	\end{tabular}
	\end{center}
\end{table}

As seen from the results in Table~\ref{tab:spectrogramcalc}, the read noise can be indeed considered white, with the non-whiteness coefficient $F\approx 1$, which is consistent with the model~\cite{boiecoxccdnoisemodel}. The non-flatness coefficient for the hardware sensor is slightly greater than that of the simulations; the uncertainty of the measurements is larger as well.

\subsection{Dark signal performance for different integration times}\label{subsec:darkfpnperf}
\begin{figure}[ht!]
\center{\includegraphics[width=0.99\linewidth]{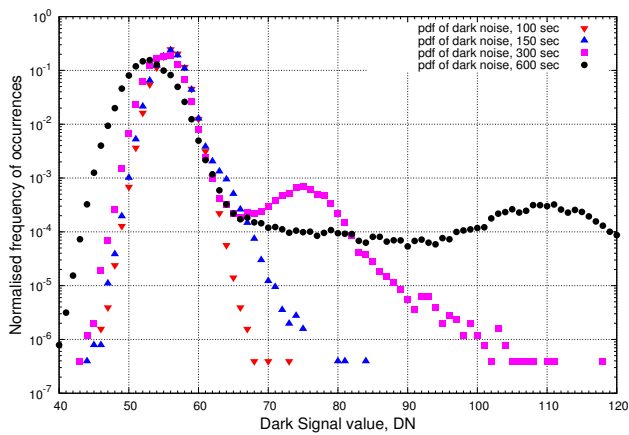}}
\caption{Normalised frequency of occurrences of dark signal values in the averaged dark image for long integration time: experimental results for 100, 150, 300 and 600 seconds.}
\label{fig:plotDarkSignalLongTimemultiple3in1_longtime_experiment}
\end{figure}

In this subsection we discuss the complexity of the dark current FPN models. It is shown that the assumptions usually made in the literature (i.e., that the dark signal is Gaussian) are not always consistent with the noise in the hardware photosensor. The experimental results of dark signal for long integration times (see Fig.~\ref{fig:plotDarkSignalLongTimemultiple3in1_longtime_experiment}) illustrate the complexity of the dark signal: while for rather short integration time times (less than 100 seconds) the distribution can be approximated as Gaussian, longer integration times have a lot more complicated structure. This is important is such applications as, e.g., astronomy, where the integration times can be very long for dim objects.

We measured the dark \label{review2JEI:R1Q6} signal\footnote{By \textit{``dark signal''} we mean the dominant component of the noise in the absence of light. Our main concern in this section is dark current Fixed Pattern Noise (dark current FPN) and comparing actual noise with various models. In case of short integration times, the dominant component of the dark signal is dark current FPN (although other components also contribute to the overall noise picture). At long integration times the noise measurements are dominated by thermally-generated electron noise, but the dark current FPN is still a considerable contributor. Since our main concern in this section is the dark current FPN and its modelling, we prefer to keep to calling it as such, even though there are other contributing sources of dark noise, especially at long integration times.}  of the hardware CMOS sensor for two different cases:
\begin{enumerate}
\item \textit{short integration time} (less than 100 seconds): the only one probability distribution (main distribution) is used for description of the noise statistics;
\item \textit{long integration time} (longer than 100 seconds): a main distribution and a superimposed distribution are used (for very long integration times the second superimposed distribution is used as well).
\end{enumerate}

We provide the analysis of dark signal statistics using superimposed distributions. The data provided in this subsection was measured at room temperature ($+25^\circ$C) as before.

\paragraph*{Method of measurements:}
We took 32 dark frames for each integration time and averaged them in order to reduce the temporal noise and \label{review2JEI:R1Q5} get the dark current FPN data. The normalised frequency of occurrences of noise values in the averaged dark image was estimated. For the estimation of statistical parameters of the dark current FPN we used a maximum likelihood estimator (MLE). All estimated statistical parameters reported are within a 95\% confidence interval if not stated otherwise. For the description of the dark current FPN (which is the main contributor of the dark noise, especially in short integration times), we use the following distributions:
\begin{itemize}
 \item Log-Normal distribution:
 $p_{LogNorm}(x;\mu, \sigma) = \frac{1}{x\sqrt{2\pi\sigma^2}}\, e^{-\frac{\left(\ln x-\mu\right)^2}{2\sigma^2}}$
 \item Gamma distribution:
$p_{Gamma}(x; a,b) = \frac{1}{b^a \Gamma(a)} x^{a-1} \exp\left[-\frac{x}{b}\right]$
\end{itemize}
As superimposed distributions, the uniform distribution and Gaussian distribution were used, as discussed below.

\subsubsection{Short integration time}\label{sec:simulations:subsubsec:shorttime}
We measured the dark current FPN for the hardware CMOS sensor and compared the results with our high-level sensor model. The 5T CMOS hardware sensor has in-built circuitry for noise cancelling that uses Correlated Double Sampling (CDS) technique. The details of the CDS algorithm, however, have not been disclosed by the manufacturer. Consequently, the noise on the images is reduced, so the comparison of the numerical models with the hardware data is approximate.

\begin{figure}[ht!]
\center{\includegraphics[width=0.99\linewidth]{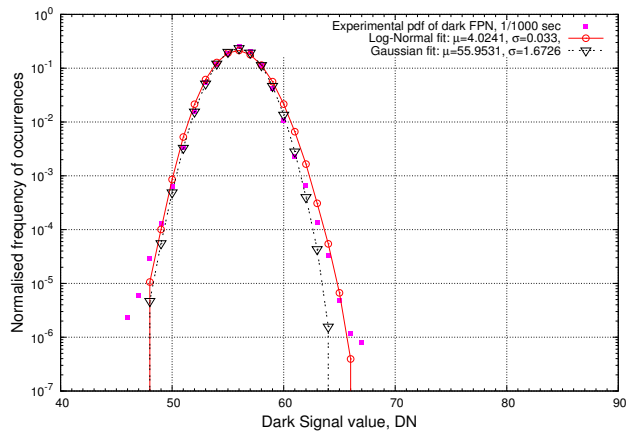}}
\caption{Normalised frequency of occurrences of dark signal values in the averaged dark image for 1/1000 seconds integration time.}
\label{fig:plotDarkSignalLongTimemultiple3in1_shorttimelinlin}
\end{figure}

The measurements presented in Fig.~\ref{fig:plotDarkSignalLongTimemultiple3in1_shorttimelinlin} were taken for integration times of 0.001 second, which are typical for, e.g., wavefront sensing in adaptive optics. The probability density function of the dark current FPN can be considered as Gaussian for the short integration times, and the ML estimated values of $\mu = 55.9531$ and $\sigma = 1.6712$  for the Gaussian distribution were obtained from the experimental data for 0.001 second. While Gaussian distribution provides and adequate description of the dark signal, it underestimates the ``outliers'' or ``dark spikes'', and this is why we use the Log-Normal distribution (see Fig.~\ref{fig:plotDarkSignalLongTimemultiple3in1_shorttimelinlin}, data marked by ``$-\circ-$ ''). The MLE parameters for the Log-Normal distribution were found as $\mu = 4.0241$ and $\sigma = 0.033$. 

\begin{figure}[ht!]
\center{\includegraphics[width=0.99\linewidth]{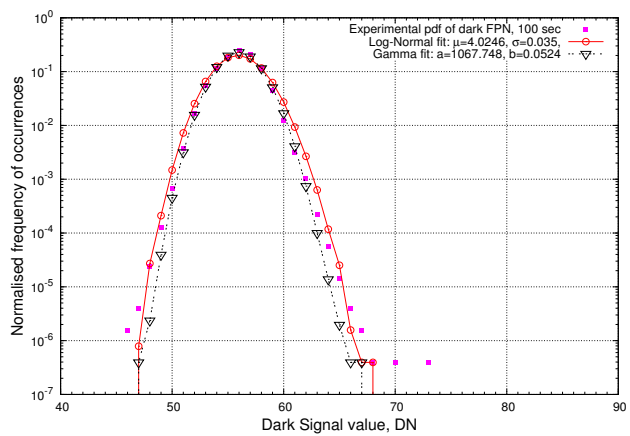} }
\caption{Normalised frequency of occurrences of noise values on the averaged dark frame for the integration time 100 seconds.}
\label{fig:plotDarkSignalLongTimemultiple3in1_shorttimesemilog100_sec_IMG_2489pdfEstimationexpVSsim}
\end{figure}

As integration time increases up to 100 seconds, the distribution of dark current FPN tends to be more asymmetric, however, the Log-Normal parameters remain similar to the very short integration time case ($\mu = 4.0246$ and $\sigma = 0.035$ for the case of 100 second integration time). Therefore, the Log-Normal distribution can still be considered as a reasonable model for the dark signal, which is consistent with the results reported in~\cite{baer2006model,dFPNinCMOS}. It is noteworthy that the Gamma distribution was found to be a better approximation for a 3T CMOS photosensor in~\cite{baer2006model}, but for our 5T CMOS sensor the Gamma distribution appears to underestimate the ``dark spikes'', as can be noted from  Fig.~\ref{fig:plotDarkSignalLongTimemultiple3in1_shorttimesemilog100_sec_IMG_2489pdfEstimationexpVSsim}.

\label{review2JEI:R1Q6-1}Also, one can notice that the dark current estimated from the measurements of the hardware sensor in this section are different from the model of dark current values in Eq.~\ref{eq:darkcurrentrate}. This is expected, since our hardware CMOS sensors has on-chip noise suppressing circuitry (see Subsection~\ref{subsec:cdsdescription}) that removes dark noise levels.

\subsubsection{Long integration time}\label{subsec:verylongfpn}
\begin{figure}[ht!]
\centering\includegraphics[width=0.99\linewidth]{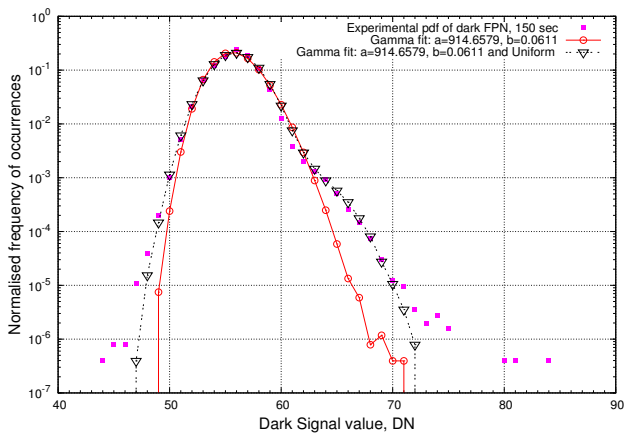}
\caption{Normalised frequency of occurrences of dark signal values in the averaged dark image for a 150 second integration time approximated by only one probability distribution (-$\circ$-) and using a superimposed probability distribution (-$\bigtriangledown$-).}
\label{fig:plotDarkSignalLongTimemultiple3in1_shorttimelinlin150_sec_IMG_0012pdfEstimationexpVSsim}
\end{figure}

The non-Gaussian nature of dark current FPN is more apparent as the integration time increases. One can see in Fig.~\ref{fig:plotDarkSignalLongTimemultiple3in1_shorttimelinlin150_sec_IMG_0012pdfEstimationexpVSsim} that either the Log-Normal or Gamma distributions alone do not describe the data adequately. Therefore, in the case of long integration time, it is necessary to superimpose another probability distribution, as mentioned in~\cite{comprehcmosmodel}, in order to create a distribution with a very long ``tail'' to  simulate ``outliers''. 

The overall distribution for the simulation of the dark signal was constructed as follows. First, the Gamma distribution was generated with parameters $a=914.6579$ and $b=0.0611$, which were estimated from the hardware CMOS photosensor data. Next, the product of two uniform distributions on the interval $[45, 440]$ was superimposed (the parameters were chosen to fit both the ``outliers'' and to widen the left part of the probability density). The results are shown in Fig.~\ref{fig:plotDarkSignalLongTimemultiple3in1_shorttimelinlin150_sec_IMG_0012pdfEstimationexpVSsim}. One can see a better agreement of the superimposed distribution described above with the experimental estimation of the dark signal statistics.

\begin{figure}[ht!]
\center{\includegraphics[width=0.99\linewidth]{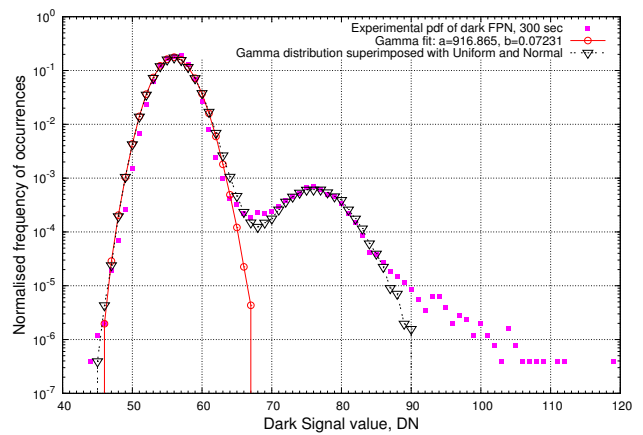}}
\caption{Normalised frequency of occurrences of dark signal values in the averaged dark image for long integration time for very long integration time (300 seconds) using a superimposed distribution.}
\label{fig:plotDarkSignalLongTimemultiple3in1_longtime}
\end{figure}

The dark signal statistics is getting more complicated with very long integration time. The dark current FPN is no longer the main contributor to the noise statistics: one would expect the noise measurements to be dominated by thermally-generated electron noise. Those are the two distributions (see Fig.~\ref{fig:plotDarkSignalLongTimemultiple3in1_longtime}) that underlie the dark signal measurements. The dark signal in these cases may require modelling based on the experimental data for a particular type of a photosensor. In our example, we were able to compose a distribution to describe the experimental dark signal probability distribution using the superimposed distributions as follows:
\begin{enumerate}
 \item main distribution: Gamma with parameters $a = 627.200$ and $b = 0.0893$;
 \item superimposed: uniform distribution on interval $[50, 460]$;
\item superimposed: Gaussian distribution with parameters $\mu = 18$ and $\sigma = 4$.
\end{enumerate}

The composed distribution describes the dark current FPN ( $\blacksquare$ symbol) with good accuracy, as seen in Fig.~\ref{fig:plotDarkSignalLongTimemultiple3in1_longtime}. One can see from the provided data that the approximation of dark current FPN using only a Gaussian distribution is poor. On the other hand, the superimposed distributions provide a better description of the complex structure of the dark current FPN.

\subsubsection{Results discussion}\label{review:2-1R1}\label{subsec:darknoisemeasurementdiscussion}
The modelling of the dark signal is complicated, since it is difficult to derive a model for the parameters of the superimposed probability distributions. The description of long-exposure dark current FPN is complicated because the statistical parameters are dependent on temperature and exposure (integration) time in a non-trivial way and likely to vary from one photosensor to another. An additional problem for the analysis of the dark signal is that the exact CDS algorithm is usually not disclosed by the manufacturer, and even if disclosed, it is difficult to simulate precisely. Furthermore, additional on-chip (e.g., transfer noise) and off-chip noise (clock-jitter noise, preamplifier noise, etc.) can contribute~\cite{janesickscintificCCD} to the resulting noise distribution. 

Nonetheless, the description based on superimposed distributions gives acceptable results for the fixed sensor's parameters (temperature, photosensor technology). The experimental data along with the models developed in this subsection describe the dark current FPN noise for different integration times. Three cases were considered:
\begin{itemize}
 \item \textit{short integration time (less than 100 seconds)}, where a LogNormal distribution is adequate as a main distribution,
 \item \textit{long integration time (longer than 100-150 seconds)}, where another distribution must be superimposed to a main distribution (or several, in case of integration times longer than 300 seconds).
\end{itemize}

At short integration times, the noise is the variability of dark current across pixels, that is, dark current FPN (or dark signal non-uniformity, DSNU).   \label{review2JEI:R2_Q15} At long integration times, one would expect the noise measurements to be dominated by thermally-generated electron noise. Those are the two distributions (see Fig.~\ref{fig:plotDarkSignalLongTimemultiple3in1_longtime_experiment} and Fig.~\ref{fig:plotDarkSignalLongTimemultiple3in1_longtime}) that presumably underlie the measurements. As the integration time increases further, the probability distribution of the resulting dark signal gets more complicated and harder to model.

\section{Conclusion}\label{sec:conclusion}
% 1. What the problem was addressed?
The article provides a literature review of the noise models that are used for the simulations of noise in CCD and CMOS photosensors. A high-level model of CCD and CMOS photosensors based on a literature review is formulated in this paper. The experimental results for a hardware 5T CMOS photosensor are presented as a validation for the developed model of a photosensor.

% 2. What are results?
The formulated model of the CMOS photosensor was compared with the data from a custom-made hardware photosensor for validation of the photosensor model. Results of Photon Transfer Curve (PTC) estimation show the importance of the non-linearity introduced in the model. Using the models of V/V and the ADC non-linearity allow to better match the properties of the hardware CMOS sensor. It has been also demonstrated that the dark current FPN and in particular the dark ``spikes'' in the case of a long integration time has a more complicated distribution than previously discussed in the literature. The Gamma distribution with a superimposed uniform and Gaussian distributions are shown to be a reasonable approximation of the dark ``spikes'' for long integration time.

% 3. So what?
The article summarises the efforts in numerical simulations of solid-state photosensors providing an extensive survey of literature on noise in photosensors. A high-level model of CMOS/CCD sensors has been formulated in order to provide engineers with insight on the noise impact on images quality. Using such a high-level photosensor model, one can simulate either CCD or CMOS sensors by application of appropriate noise models. The formulated high-level simulation model can be used to synthesise realistically noised images for the development and testing of the image processing algorithms.

\section*{Acknowledgements}
The authors are gratefully acknowledge personal communications with \textit{Dr. James R. Janesick} and his valuable comments.

\small
% \bibliographystyle{unsrt}
% \bibliography{my}

\end{document}